\documentclass[suppldata]{interact}

\usepackage[natbibapa,nodoi]{apacite}
\theoremstyle{plain}

\theoremstyle{definition}

\theoremstyle{remark}

\usepackage{indentfirst}
\usepackage[utf8]{inputenc}
\usepackage{amsmath}
\usepackage{mathrsfs}
\usepackage{amssymb}
\usepackage{color}
\usepackage{comment}
\usepackage{float}
\usepackage{graphicx}
\usepackage{bernsteinStyle}
\usepackage{float}
\usepackage{tikz}
\usepackage{booktabs}
\usepackage{pgfplots}
\usepackage{array}
\usepackage{bm}
\usepackage{makecell}
\usepackage{multirow}
\usepackage[hidelinks]{hyperref}
\newcolumntype{C}[1]{>{\centering\let\newline\\\arraybackslash\hspace{0pt}}m{#1}}

\usepackage{algorithm}
\usepackage{algpseudocode}
\usepackage{caption}
\newcommand{\algindent}{\hspace{\algorithmicindent}}

\usepackage{pgfplots}
\pgfplotsset{compat=1.18}

\makeatother

\title{\large Real-Time Numerical Differentiation of Sampled Data Using Adaptive Input and State Estimation} 
\author{
\name{Shashank~Verma\textsuperscript{a}\thanks{CONTACT Shashank Verma. Email: shaaero@umich.edu}, Sneha~Sanjeevini\textsuperscript{a}\thanks{CONTACT Sneha Sanjeevini. Email: snehasnj@umich.edu}, E.~Dogan~Sumer,\textsuperscript{b}\thanks{CONTACT E. Dogan Sumer. Email: esumer@ford.com} and Dennis S. Bernstein\textsuperscript{a}\thanks{CONTACT Dennis S. Bernstein. Email: dsbaero@umich.edu}}
\affil{\textsuperscript{a}Department of Aerospace Engineering, University of Michigan, Ann Arbor, Michigan, USA
\textsuperscript{b}Ford Motor Company, Dearborn, Michigan, USA}
}

\begin{document}
\maketitle
\graphicspath{
{PaperFigures/BiographyPhotos}
{PaperFigures/Numerical_Exam}
{PaperFigures/Adap_Input_State_Est}
{PaperFigures/NumDiffSamData}
{PaperFigures/Application_Carsim}
}

\begin{abstract}
Real-time numerical differentiation plays a crucial role in many digital control algorithms, such as PID control, which requires numerical differentiation to implement derivative action.  
This paper addresses the problem of numerical differentiation for real-time implementation with minimal prior information about the signal and noise using adaptive input and state estimation. 
Adaptive input estimation with adaptive state estimation (AIE/ASE) is based on retrospective cost input estimation, while adaptive state estimation is based on an adaptive Kalman filter in which the input-estimation error covariance and the measurement-noise covariance are updated online. 
The accuracy of AIE/ASE is compared numerically to several conventional numerical differentiation methods.  Finally, AIE/ASE is applied to simulated vehicle position data generated from CarSim.
\end{abstract}


\section{Introduction}

The dual operations of integration and differentiation provide the foundation for much of mathematics.
From an analytical point of view, differentiation is simpler than integration;  consider the relative difficulty of differentiating and integrating $\log(1+\sin^2 x^3).$
In numerical analysis, integration techniques have been extensively developed in \citet{davis}, whereas differentiation techniques have been developed more sporadically in \citet{savitzky1964smoothing,cullum}, \citet[pp. 565, 566]{hamming}.

In practice, numerical integration and differentiation techniques are applied to sequences of measurements, that is, discrete-time signals composed of sampled data.
Although, strictly speaking, integration and differentiation are not defined for discrete-time signals, the goal is to compute a discrete-time ``integral'' or ``derivative'' that approximates the true integral or derivative of the pre-sampled, analog signal.

In addition to sampling, numerical integration and differentiation methods must address the effect of sensor noise.
For numerical integration, constant noise, that is, bias, leads to a spurious ramp, while stochastic noise leads to random-walk divergence.
Mitigation of these effects is of extreme importance in applications such as inertial navigation as shown in \citet{farrellAN,grewal}.

Compared to numerical integration, the effect of noise on numerical differentiation is far more severe.
This situation is due to the fact that, whereas integration is a bounded operator on a complete inner-product space, differentiation is an unbounded operator on a dense subspace.
Unboundedness implies a lack of continuity, which is manifested as high sensitivity to sensor noise.
Consequently, numerical differentiation typically involves assumptions on the smoothness of the signal and spectrum of the noise as considered in \citet{ramm2,Jauberteau2009,Stickel2010,zhao2013,knowles_methods,haimovich2022}.

Numerical differentiation algorithms are crucial elements of many digital control algorithms.
For example, PID control requires numerical differentiation to implement derivative action as presented in \citet{PID2012,PIDastrom}.
%
Flatness-based control is based on a finite number of derivatives as shown in \citet{Nieuwstadt1998,mboupnumeralg}.

In practice, analog or digital filters are used to suppress the effect of sensor noise, thereby allowing the use of differencing formulae in the form of inverted ``V'' filters, which have the required gain and phase lead at low frequencies and roll off at high frequencies.
These techniques assume that the characteristics of the signal and noise are known, thereby allowing the user to tweak the filter parameters.
When the signal and noise have unknown and possibly changing characteristics, filter tuning is not possible and thus the problem becomes significantly more challenging.
The recent work in \citet{Kutz2020} articulates these challenges and proposes a Pareto-tradeoff technique for addressing the absence of prior information.
Additional techniques include high-gain observer methods, in which the observer approximates the dynamics of a differentiator 
as shown in \citet{dabroom_discrete-time_1999}.  
An additional approach to this problem is to apply
sliding-mode algorithms as shown in \citet{arie2003slidemode,reichhartinger_arbitrary-order_2018,lopez-caamal_generalised_2019,mojallizadeh2021,alwi_adap_sliding_mode_2012}.

In feedback control applications, which require real-time implementation, phase shift and latency in numerical differentiation can lead to performance degradation and possibly instability.
Phase shift arises from filtering, whereas latency arises from noncausal numerical differentiation, that is, numerical differentiation algorithms that require future data.
For real-time applications, a noncausal differentiation algorithm can be implemented causally by delaying the computation until the required data are available.
A feedback controller requires an estimate of the {\it current} derivative, however, and thus the delayed estimate provided by a noncausal differentiation algorithm may not be sufficiently accurate.

A popular approach to numerical differentiation is to apply state estimation with integrator dynamics, where the state estimate includes an estimate of the derivative of the measurement as shown in \citet{kalata_1984,bogler_1987}.
This approach has been widely used for target and vehicle tracking  
in \citet{jia_2008,karsaz_2009,lee_1999,rana_2020}.  
As an extension of this approach, the present paper applies input estimation to numerical differentiation, where the goal is to estimate the input as well as the state, as in
\citet{gillijns2007unbiased,orjuela2009simultaneous,fang2011stable,yong2016unified,hsieh2017unbiased,naderi2019unbiased,ZakTAC2021}.

The present paper is motivated by the situation where minimal prior information about the signal and noise is  available.
This case arises when the spectrum of the signal changes slowly or abruptly in an unknown way, and when the noise characteristics vary due to changes in the environment, such as weather.
With this motivation, 
adaptive input estimation (AIE) was applied to target tracking in \citet{ansari_input_2019}, where it was used to estimate  vehicle acceleration using position data.
In particular, the approach of \citet{ansari_input_2019} is based on retrospective cost input estimation (RCIE), where recursive least squares (RLS) is used to update the coefficients of the estimation subsystem.
The error metric used for adaptation is the residual (innovations) of the state estimation algorithm, that is, the Kalman filter.
This technique requires specification of the covariances of the process noise, input-estimation error, and sensor noise.

The present paper extends the approach of \citet{ansari_input_2019} by replacing the Kalman filter with an adaptive Kalman filter in which the input-estimation error and sensor-noise covariances are updated online.
Adaptive extensions of the Kalman filter to the case where the variance of the disturbance is unknown are considered in \citet{Yaesh2008SimplifiedAE,shiadapukf2009,moghe2019adaptivekfLTI,zhangadapKF2020}.
Adaptive Kalman filters based on the residual for integrating INS/GPS systems are discussed in \citet{Mohamed1999,Hide2003,Almagbile2010}. 
Several approaches to adaptive filtering, such as bayesian, maximum likelihood, correlation, and covariance matching, are studied in \citet{Mehra1972}. 
A related algorithm involving a covariance constraint is developed in \citet{junkins1988minimum}.

The adaptive Kalman filter used in the present paper as part of AIE/ASE is based on a search over the range of input-estimation error covariance.
This technique has proven to be easy to implement and effective in the presence of unknown signal and noise characteristics.
The main contribution of the present paper is { a numerical investigation of the accuracy of AIE combined with   the proposed adaptive state estimation (ASE) in the presence of noise with unknown properties.}
The accuracy of AIE/ASE is compared to the backward-difference differentiation, Savitzky-Golay differentiation (\citet{savitzky1964smoothing,SG_lecture_notes_Schafer,SG_staggs,mboupnumeralg}), and numerical differentiation based on high-gain observers (\citet{dabroom_discrete-time_1999}).

The present paper represents a substantial extension of preliminary results presented in \citet{shashankACC2022}.
In particular, the algorithms presented in the present paper extend the adaptive estimation component of the approach of \citet{shashankACC2022} { in Section \ref{sec:AdapInptStateEst}}, and the accuracy of these algorithms is more extensively evaluated and compared to prior methods { in Section \ref{sec:Numdiff_HS}.}

The contents of the paper are as follows.
Section \ref{sec:NumDiffSamData} presents three baseline numerical differentiation algorithms. 
Section \ref{sec:CompDelayAlgo} discusses the delay in the availability of the estimated derivative due to the computation time and non-causality. 
This section also defines an error metric for comparing the accuracy of the algorithms considered in this paper.
Section \ref{sec:AdapInptEst} describes the adaptive input estimation algorithm. 
Section \ref{sec:AdapInptStateEst} provides the paper's main contribution, namely, adaptive input estimation with adaptive state estimation.
Section \ref{sec:Numdiff_HS} applies three variations of AIE using harmonic signals with various noise levels.  
Finally, Section \ref{sec:AppCarsim} applies the variations of AIE to simulated vehicle position data generated by CarSim.

\section{Baseline Numerical Differentiation Algorithms} \label{sec:NumDiffSamData}

This section presents three algorithms for numerically  differentiating sampled data.
These algorithms provide a baseline for evaluating the accuracy of the adaptive input and state estimation algorithms described in Section \ref{sec:AdapInptStateEst}.

{\subsection{Problem Statement}}

Let $y$ be a continuous-time signal with $q$th derivative $y^{(q)}.$
We assume that the sampled values  $y_k\isdef y(kT_\rms)$ are available, where $T_\rms$ is the sample time.
The goal is to use the sampled values $y_k$ to obtain an estimate $\hat y^{(q)}_k$ of $y^{(q)}_k\isdef y^{(q)}(kT_\rms)$ {in the presence of the noise with unknown properties.} 
This paper focuses on the cases $q=1$ and $q=2$. In later sections, the sample values will be corrupted by noise.

\subsection{ Backward-Difference (BD) Differentiation} \label{subsec:Diff_BD}

Let $\bfq^{-1}$ denote the backward-shift operator. 
Then the {\it backward-difference single differentiator} is given by 
\begin{align}
    G_{\rm sd}(\bfq^{-1}) \isdef \cfrac{1-\bfq^{-1}}{T_\rms},
\end{align}
and the {\it backward-difference double differentiator} is given by
\begin{align}
    G_{\rm dd}(\bfq^{-1}) \isdef \cfrac{(1-\bfq^{-1})^2}{(T_\rms)^2}.
\end{align}
%

\subsection{ Savitzky–Golay (SG) Differentiation \label{subsec:Diff_SG}}

%
As shown in \citet{savitzky1964smoothing,SG_lecture_notes_Schafer,SG_staggs}, 
in SG differentiation at each step $k,$ a polynomial  
\begin{align} 
  P_k(s)  = \sum_{i=0}^{p_{\rmd}} a_{i,k} s^{i} \label{Pdefn}
  \end{align}
of degree $p_\rmd$ is fit over a sliding data window of size $2\ell+1$ centered at step $k$, where $\ell\ge 1.$ 
At each step $k$, this leads to the least-squares problem
\begin{align}
  \min\Vert\mathcal{Y}_k - \mathcal{A}_k\mathcal{X}_k\Vert, \label{sg_linearsystem}
\end{align}
where
\begin{align}
  \mathcal{Y}_k &\isdef
  \begin{bmatrix}
  y_{k{-}\ell}\\
  \vdots\\
  y_{k{+}\ell}
  \end{bmatrix},\quad 
  \mathcal{X}_k \isdef
  \begin{bmatrix}
  a_{0,k}\\
  \vdots\\
  a_{p_{\rmd},k}
  \end{bmatrix},
\end{align}\label{} 
\begin{align}
\mathcal{A}_k &\isdef
  \begin{bmatrix}
  1 & {(k-\ell)T_\rms} & ... & ({(k-\ell)T_\rms})^{p_{\rmd}}\\
  \vdots & \vdots & \ddots & \vdots\\
  1 & {(k+\ell)T_\rms} & ... & ({(k+\ell)T_\rms})^{p_{\rmd}}\\
  \end{bmatrix}.
 \end{align}
Solving \eqref{sg_linearsystem} with $q \le p_\rmd \le 2\ell$ yields 
\begin{equation}
    \hat{\mathcal{X}}_k =
  \begin{bmatrix}
  \hat a_{0,k}\\
  \vdots\\
  \hat a_{p_{\rmd},k}
  \end{bmatrix}.
\end{equation}
Differentiating \eqref{Pdefn} $q$ times with respect to $s$, setting $s=kT_\rms,$ and replacing the coefficients of $P_k$ in \eqref{Pdefn} with the components of $\hat{\mathcal{X}}_k$, the estimate $\hat{y}^{(q)}_k $ of $y^{(q)}_k$  is given by
\begin{equation} 
  \hat{y}^{(q)}_k  = \sum_{i=q}^{p_{\rmd}} Q_{i,q}\hat{a}_{i,k} (k T_\rms)^{i-q}, \label{sg_nd_est}
\end{equation}
%
where, for all $i=q,\ldots, p_\rmd,$ 
\begin{equation}
Q_{i,q}\isdef\prod_{j=1}^{q}(i-j+1).
\end{equation}
%

  \subsection{High-Gain-Observer (HGO) Differentiation   \label{subsec:Diff_HGO}}
%
A state space model for the $r$th-order continuous-time HGO is given by
\begin{align}
    \dot{\hat{x}} &= A_{\rmc\rmo} \hat{x} + B_{\rmc\rmo} y, \quad
    \hat{y} = C_\rmo \hat{x}, \label{obs_ct} \\ 
    A_{\rmc\rmo} &\isdef \begin{bmatrix}
      0_{(r-1) \times 1} & I_{r-1} \\
      0 & 0_{1\times (r-1)}
  \end{bmatrix}-H\begin{bmatrix}
    1 & 0_{1\times (r-1)}
\end{bmatrix},\label{hgo_Aco}\\
 C_\rmo &\isdef \begin{bmatrix} 0_{(r-1)\times 1} & I_{r-1} \end{bmatrix}, \label{hgo_Cco}\\
    B_{\rmc\rmo} &= H \isdef \begin{bmatrix}
        \cfrac{\alpha_1}{\varepsilon} & \cfrac{\alpha_2}{\varepsilon^2} & \cdots &  \cfrac{\alpha_r}{\varepsilon^r} \label{hgo_gain}
    \end{bmatrix}^\rmT, 
\end{align}
where $\varepsilon>0$  and $\alpha_1,\ldots,\alpha_r$ are constants chosen such that the polynomial
\begin{align}
p(s) \isdef s^r + \alpha_1 s^{r-1} + \cdots + \alpha_{r-1} s + \alpha_r
\end{align}
is Hurwitz. 
The transfer function from $y$ to $\hat{y}$ is given by
\begin{align}
G(s) &= C_\rmo(sI - A_{\rmc\rmo})^{-1} H = D_{G}^{-1}(s)N_{G}(s),
\end{align}
where
\begin{align}
\hspace{-0.2cm}D_G(s) &\isdef \varepsilon^rs^r + \alpha_1\varepsilon ^{r-1}s^{r-1} + \dots + \alpha_{r-1}\varepsilon s + \alpha_r, 
\end{align}
\begin{align}
N_G(s) &\isdef \begin{bmatrix}
\alpha_2\varepsilon ^{r-2}s^{r-1} + \dots + \alpha_{r-1}\varepsilon s^2 + \alpha_r s\\
         \alpha_3\varepsilon ^{r-3}s^{r-1} + \dots + \alpha_{r-1}\varepsilon s^3 + \alpha_rs^2 \\
         \vdots \\
         \alpha_{r-1}\varepsilon s^{r-1} + \alpha_rs^{r-2} \\
         \alpha_rs^{r-1}
         \end{bmatrix}.
\end{align}
Since 
\begin{align}
   \displaystyle{\lim_{\varepsilon \to 0}} G(s) = \begin{bmatrix}
       s & \cdots  & s^{r-1}
   \end{bmatrix}^\rmT, \label{lim_tf}
\end{align}
it follows that, for all $i=1, \dots, r-1$, the 
$i$th component of $\hat{y}$ is an approximation of $y^{(i)}$.
Applying the bilinear transformation to \eqref{obs_ct} yields the discrete-time observer 
\begin{align}
    \hat{x}_{k+1} = A_{\rmd\rmo} \hat{x}_k + B_{\rmd\rmo} y_k, \quad
    \hat{y}_{k} = C_\rmo \hat{x}_k, \label{obs_dt}
\end{align}
where
\begin{align}
  A_{\rmd\rmo} &\isdef (I_{\rmr} - \half T_\rms A_{\rmc\rmo} )^{-1}(I_{\rmr} + \half T_\rms A_{\rmc\rmo} ),\label{hgo_Ado} \\
  B_{\rmd\rmo} &\isdef (I_{\rmr} - \half T_\rms A_{\rmc\rmo} )^{-1}B_{\rmc\rmo}T_\rms.\label{hgo_Bdo}
\end{align}
Implementation of \eqref{obs_dt} provides estimates $\hat{y}^{(1)}_k,  \dots, \hat{y}^{(r-1)}_k$ of $y^{(1)}_k, \dots, y^{(r-1)}_k$.
%


\section{Real-Time Implementation and Comparison of Baseline Algorithms}  
\begin{figure}[H]
  \begin{center}
  {\includegraphics[width=0.75\columnwidth]{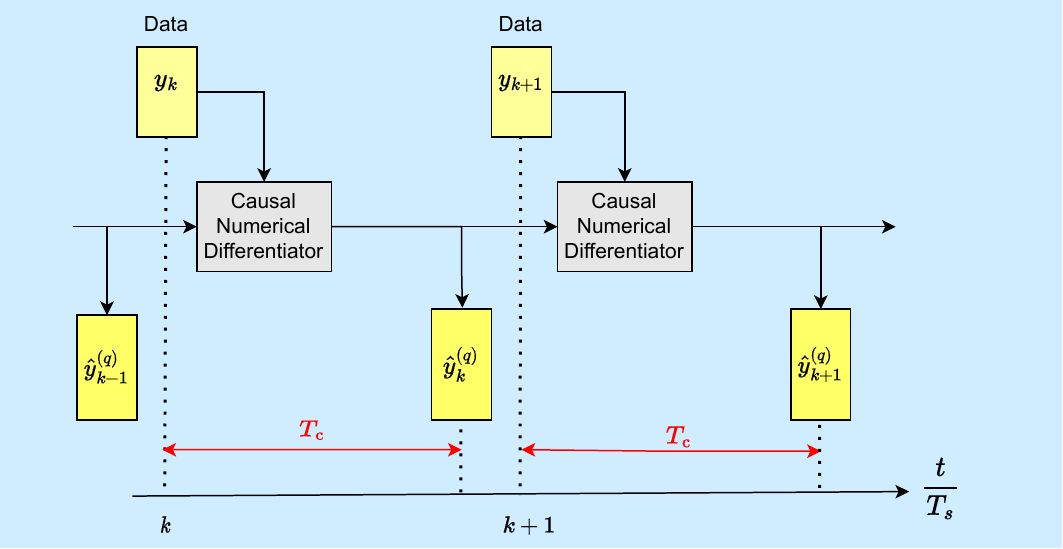}}
  \end{center}
  \caption{\textit{Timing diagram for causal numerical differentiation.} The causal numerical differentiator uses data obtained at step $k$ to estimate the derivative of the signal $y$.
  Because of the computation time $T_\rmc,$ the estimate $\hat y_k^{(q)}$ of $y_k^{(q)}$ is not available until step $k+1$. 
 In this case, the delay is $\delta = 1$ step. For a noncausal differentiator, $\delta \ge 2.$}
  \label{fig:sec3_general_timing_diag}
\end{figure}

\textbf{\label{sec:CompDelayAlgo}}
{
Several noteworthy differences exist among BD, SG, and HGO. First,  BD differentiation  operates on adjacent pairs of data points, whereas SG differentiation operates on a moving window of data points. Consequently,  SG differentiation is potentially more accurate than BD differentiation.}
Because of the time $T_\rmc$ needed for computation,  all numerical differentiation entails an unavoidable delay in the availability of the estimated derivative.
In addition, BD differentiation and HGO differentiation do not require future data to estimate the derivative, thus they are causal algorithms. 
On the other hand, SG differentiation requires future data, and thus it is noncausal.
For causal differentiation, the delay is $\delta = 1$  step due to computation; for noncausal differentiation, $\delta \ge 2$.
Assuming $T_\rmc \le T_\rms,$ Figure \ref{fig:sec3_general_timing_diag} illustrates the case $\delta=1$. 
Note that $\delta = 1$ for BD and HGO, whereas $\delta = \ell + 1$ for SG with window size $2\ell +1.$

To quantify the accuracy of each numerical differentiation algorithm, for all $k\ge \delta$, we define the relative root-mean-square error (RMSE) of the estimate of the $q$th derivative as
%
%
\begin{align}
  \rho_k^{(q)} \isdef
\sqrt{\frac{\displaystyle\sum_{i=\delta}^{k}({y}_{i}^{(q)}-\hat y_{i-\delta}^{(q)})^2}{\displaystyle\sum_{i=\delta}^{k} (y_{i-\delta}^{(q)})^2}}.  \label{rms}
\end{align} 
Note that numerator of \eqref{rms} accounts for the effect of the delay $\delta.$
{  For real-time implementation, the relevant error metric depends on the difference between the true current derivative and the currently available estimate of the past derivative, as can be seen in the numerator of \eqref{rms}. When the derivative estimates are exact, \eqref{rms} determines an RMSE value that can be viewed as the {\it delay floor} for the $q$th derivative, that is, the error due solely to the fact that a noncausal differentiation algorithm must be implemented with a suitable delay.
}
Note that the delay floor depends on $\delta$ and is typically positive.

The {\it true} values of $y^{(q)}_k$ are the sampled values of $y^{(q)}$ in the absence of sensor noise.
Of course, the true values of $y^{(q)}_k$ are unknown in practice and thus cannot be used as an online error criterion.
However, these values are used in \eqref{rms}, which is computable in simulation for comparing the accuracy of the numerical differentiation algorithms.

To compare the various baseline algorithms presented in Section \ref{sec:NumDiffSamData}, we consider numerical differentiation of the continuous-time signal $y(t) = \sin(20t)$, where $t$ is time in seconds.
The signal $y(t)$ is sampled with sample time $T_\rms = 0.01$ sec. 
The measurements are assumed to be corrupted by noise, and thus the noisy sampled signal is given by $y_k = \sin(0.2k)+Dv_k$, where $v_k$ is standard (zero-mean, unit-variance, Gaussian) white noise.
The value of $D$ is chosen to set the desired signal-to-noise ratio (SNR).

\begin{figure}[H]

    \begin{center}
  {\includegraphics[width=0.75\linewidth]{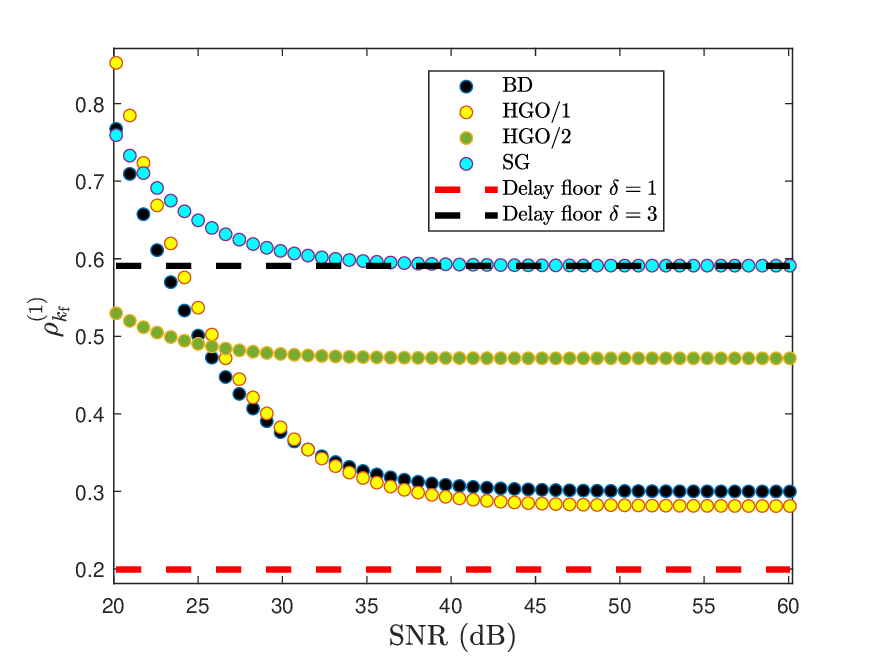}}
  \end{center}
    \caption{Relative RMSE $\rho_{k_\rmf}^{(1)}$ of the estimate of the first derivative versus SNR, where $k_{\rmf} = 2000$ steps, for BD, SG, HGO/1, and HGO/2.
    For the first derivative, the red dashed line denotes the delay floor for $\delta=1$, and the black dashed line denotes the delay floor for $\delta=3$.
    %
    }
    \label{fig:sec2_RMSE_SNR_SD}
  \end{figure}

For single differentiation with SG, let $\ell = 2$ and $p_{d} = 3$.
For single differentiation with HGO, 
let HGO/1 denote HGO with
$r = 2$, $\alpha_1 = 2$, $\alpha_2 = 1$, and $\varepsilon = 0.2$,
and let HGO/2 denote HGO/1 with $\varepsilon = 0.2$ replaced by $\varepsilon = 0.7$.
Figure \ref{fig:sec2_RMSE_SNR_SD} shows the relative RMSE $\rho_{k_\rmf}^{(1)}$ of the estimate of the first derivative for SNR ranging from $20$ dB to $60$ dB, where $k_\rmf = 2000$ steps.

\begin{figure}[H]
    
      \begin{center}
    {\includegraphics[width=0.75\linewidth]{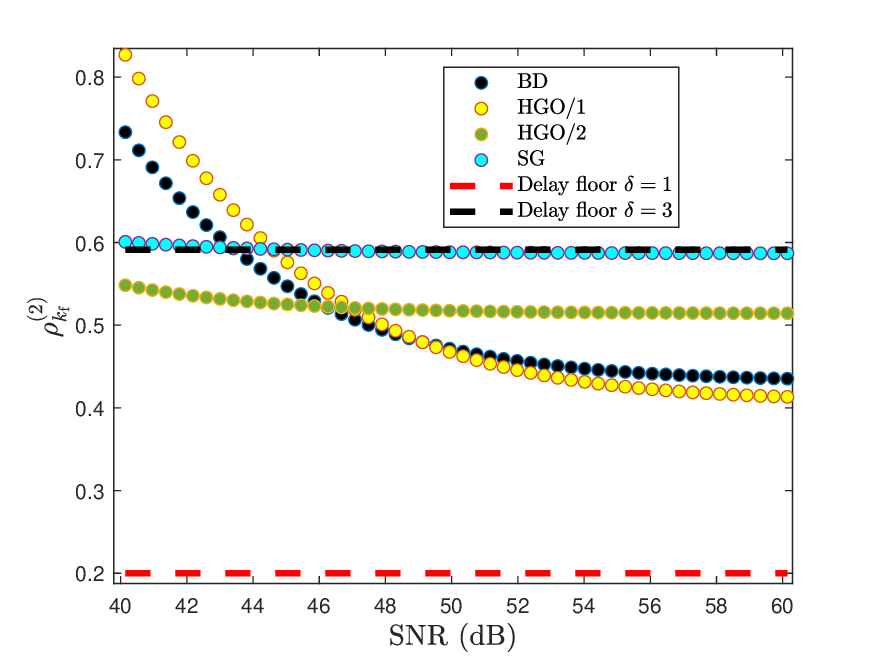}}
    \end{center}
    \caption{Relative RMSE $\rho_{k_\rmf}^{(2)}$ of the estimate of the second derivative versus SNR, where $k_{\rmf} = 2000$ steps, for BD, SG, HGO/1, and HGO/2. 
    For the second derivative, the red dashed line denotes the delay floor for $\delta=1$, and the black dashed line denotes the delay floor for $\delta=3$.
    }
    \label{fig:sec2_RMSE_SNR_DD}
    \end{figure}

For double differentiation with SG, let $\ell = 2$ and $p_{d} = 2$.
For double differentiation with HGO, 
let HGO/1 denote HGO with
$r = 3$, $\alpha_1 = 8$, $\alpha_2 = 24$, $\alpha_3 = 32$, and $\varepsilon = 1$,
and let HGO/2 denote HGO/1 with $\varepsilon = 1$ replaced by $\varepsilon = 2$.
Figure \ref{fig:sec2_RMSE_SNR_DD} shows the relative RMSE $\rho_{k_\rmf}^{(2)}$ of the estimate of the second derivative for SNR ranging from $40$ dB to $60$ dB, where $k_\rmf = 2000$ steps.
{The delay floors in Figures \ref{fig:sec2_RMSE_SNR_SD} and 
 \ref{fig:sec2_RMSE_SNR_DD} are determined by computing the RMSE between the true value $y_{k}^{(q)}$ and the $\delta$-step-delayed true value $y_{k-\delta}^{(q)}$. }

{
The comparison between HGO/1 and HGO/2 in Figure \ref{fig:sec2_RMSE_SNR_SD} and \ref{fig:sec2_RMSE_SNR_DD} shows that the performance of HGO differentiation depends on the noise level, and thus tuning is needed to achieve the best possible performance.
When the noise level is unknown, however, this tuning is not possible.
Hence, we now consider a differentiation technique that adapts to the actual noise characteristics.
}

\section{Adaptive Input Estimation} \label{sec:AdapInptEst}

This section discusses adaptive input estimation (AIE), which is a specialization of retrospective cost input estimation (RCIE) derived in \citet{ansari_input_2019}. 
This section explains how AIE specializes RCIE to the problem of causal numerical differentiation.

Consider the linear discrete-time system
\begin{align}
	x_{k+1} &=  A x_{k} + Bd_{k}, 	\label{state_eqn}\\
	y_k  &= C x_k + D_{2} v_k, \label{output_eqn}
\end{align}
where
$k\ge0$ is the step,
$x_k \in \mathbb R^{n}$ is the state,
$d_k \isdef d(kT_\rms) \in \mathbb R$,
$v_k \in \mathbb R$ is standard white noise, and $D_2v_k \in \mathbb R$ is the sensor noise.
The matrices $A \in \mathbb R^{n \times n}$, $B \in \mathbb R^{n \times 1}$, $C \in \mathbb R^{1 \times n}$, and $D_{2} \in \mathbb R$ are assumed to be known.
Define the sensor-noise covariance $V_{2} \isdef D_{2} D_{2}^\rmT$.
The {\it goal} of AIE is to estimate $d_k$ and $x_k$.
%

%
AIE consists of three subsystems, namely, the Kalman filter forecast subsystem, the input-estimation subsystem, and the Kalman filter data-assimilation subsystem.
First, consider the Kalman filter forecast step
\begin{gather}
	x_{{\rm fc},k+1} = A x_{{\rm da},k} + B \hat{d}_{k},	\label{kalman_fc_state}\\
	y_{{\rm fc},k} =  C x_{{\rm fc},k}, \label{kalman_fc_output}\\
	z_k = y_{{\rm fc},k} - y_k, 		\label{innov_error}
\end{gather}
where
$\hat d_k$ is the estimate of $d_k$, 
$x_{\rm da,k} \in \mathbb R^{n}$ is the data-assimilation state, 
$x_{{\rm fc},k} \in \mathbb R^{n}$ is the forecast state,
$z_k \in \mathbb R$ is the residual, and $x_{{\rm fc},0} = 0$.

Next, to obtain $\hat{d}_k$, the input-estimation subsystem of order $n_\rme$ is given by the exactly proper dynamics
\begin{align}
\hat{d}_k = \sum\limits_{i=1}^{n_\rme} P_{i,k} \hat{d}_{k-i} + \sum\limits_{i=0}^{n_\rme} Q_{i,k} z_{k-i}, \label{estimate_law1}
\end{align}
%
where $P_{i,k} \in \BBR$ and $Q_{i,k} \in \BBR$.
AIE minimizes $z_{k}$ by using recursive least squares (RLS) to update $P_{i,k}$ and $Q_{i,k}$ as shown below.
The subsystem \eqref{estimate_law1} can be reformulated as
\begin{align}
\hat{d}_k=\Phi_k \theta_k, \label{estimate_law12}
\end{align}
where the regressor matrix $\Phi_k$ is defined by
\begin{align}
	\hspace{-0.2cm}\Phi_k \isdef
		\begin{bmatrix}
			\hat{d}_{k-1} &
			\cdots &
			\hat{d}_{k-n_{\rme}} &
			z_k &
			\cdots &
			z_{k-n_{\rme}}
		\end{bmatrix}
		\in \mathbb{R}^{1 \times l_{\theta}},
\end{align}
the coefficient vector $\theta_k$ is defined by
\begin{align}
\hspace{-0.2cm}\theta_k \isdef \begin{bmatrix}
P_{1,k} & \cdots & P_{n_{\rme},k} & Q_{0,k} & \cdots & Q_{n_{\rme},k}
\end{bmatrix}^{\rmT}
\in \mathbb{R}^{l_{\theta}},
\end{align}
and $l_\theta \isdef  2n_{\rme} +1$.
%
%
In terms of the backward-shift operator $\bfq^{-1}$, \eqref{estimate_law1} can be written as
\begin{align}
   \hat{d}_{k} = G_{\hat{d}z,k}(\bfq^{-1})z_k,
\end{align}
where
\begin{align}
    G_{\hat{d}z,k} &\isdef D_{\hat{d}z, k}^{-1}  \it{N}_{\hat{d}z,k}, \label{d_hat_z_tf} \\
    D_{\hat{d}z,k}(\bfq^{-1}) &\isdef I_{l_d}-P_{1,k}\bfq^{-1} - \cdots-P_{n_\rme,k}\bfq^{-n_\rme}, \label{d_hat_z_tf_D} \\
    N_{\hat{d}z, k}(\bfq^{-1}) &\isdef Q_{0,k} + Q_{1,k} \bfq^{-1}+\cdots+Q_{n_\rme,k}\bfq^{-n_\rme}. \label{d_hat_z_tf_N}
\end{align}

To update the coefficient vector $\theta_k,$ we define the filtered signals
\begin{align}
\Phi_{{\rm f},k} &\isdef G_{{\rm f}, k}(\bfq^{-1}) \Phi_{k}, \quad
\hat{d}_{{\rm f},k} \isdef G_{{\rm f}, k}(\bfq^{-1}) \hat{d}_{k}, \label{eq:filtdhat}
\end{align}
%
%
where, for all $k\ge 0$,
\begin{align}
G_{{\rm f}, k}(\bfq^{-1}) = \sum\limits_{i=1}^{n_{\rm f}} \bfq^{-i}H_{i,k}, \label{Gf}
\end{align}
\begin{align}
H_{i,k} &\isdef \left\{
\begin{array}{ll}
C B, & k\ge i=1,\\
C \overline{A}_{k-1}\cdots \overline{A}_{k-(i-1)}  B, & k\ge i \ge 2, \\
0, & i>k,
\end{array}
\right. 
\end{align}
and $\overline{A}_k \isdef A(I + K_{{\rm da},k}C)$, where $K_{{\rm da},k}$ is the Kalman filter gain given by \eqref{kalman_gain} below.
%
%
%
Furthermore, define the {\it retrospective variable}
\begin{align}
z_{{\rm r},k}(\hat{\theta}) \isdef z_k -( \hat{d}_{{\rm f},k} - \Phi_{{\rm f},k}\hat{\theta} ), \label{eq:RetrPerfVar} 
\end{align}
where the coefficient vector $\hat{\theta} \in \BBR^{l_\theta}$ denotes a variable for optimization,
and define the retrospective cost function
\begin{align}
\SJ_k(\hat{\theta}) \isdef \sum\limits_{i=0}^k  [R_z z_{{\rm r},i}^{2}(\hat{\theta}) +  R_{d} (\Phi_i\hat{\theta})^2] +  (\hat{\theta} - \theta_0)^\rmT R_{\theta} (\hat{\theta} - \theta_0), \label{costf}
\end{align}  
where $R_z\in(0,\infty)$, $R_d\in(0,\infty),$ and $R_{\theta}\in\BBR^{l_{\theta} \times l_{\theta}}$ is positive definite.
Then, for all $k\ge 0$, the unique global minimizer $\theta_{k+1}$ of \eqref{costf} is given by the RLS update as shown in \citet{islam2019recursive}
\begin{align}
P_{k+1} &= P_{k} - P_{k} \widetilde{\Phi}^{\rmT}_{k} \Gamma_{k} \widetilde{\Phi}_{k} P_{k}, \label{covariance_update} \\
\theta_{k+1} &= \theta_{k} - P_{k} \widetilde{\Phi}^{\rmT}_{k} \Gamma_{k}(\widetilde{z}_{k} + \widetilde{\Phi}_{k} \theta_{k}), \label{theta_update}
\end{align}
where
\begin{gather*}
P_0 \isdef R_{\theta}^{-1},\quad 
\Gamma_k \isdef ( \widetilde{R}^{-1} + \widetilde{\Phi}_k P_{k} \widetilde{\Phi}^{\rmT}_k)^{-1}, \quad
\widetilde{\Phi}_k \isdef \begin{bmatrix}
   \Phi_{\rmf, k}  \\
   \Phi_k   \\
\end{bmatrix}, \\
\widetilde{z}_k \isdef \begin{bmatrix}
   z_k-\hat{d}_{{\rm f},k}  \\
   0   \\
\end{bmatrix}, \quad
\widetilde{R} \isdef \begin{bmatrix}
   R_z & 0  \\
   0 & R_{d}   \\
\end{bmatrix}.
\end{gather*}
Using the updated coefficient vector given by \eqref{theta_update}, the estimated input at step $k+1$ is given by replacing $k$ by $k+1$ in \eqref{estimate_law12}. We choose $\theta_0 = 0,$ and thus $\hat{d}_0 = 0.$
Implementation of AIE requires that the user specify the orders $n_\rme$ and $n_\rmf,$ as well as the weightings $R_z,$ $R_d,$ and $R_\theta.$  
These parameters are specified for each example in the paper.
 
\subsection{State Estimation} 

The forecast variable $x_{{\rm fc},k}$ given by \eqref{kalman_fc_state} is used to obtain the estimate $x_{{\rm da},k}$ of $x_k$ given by the Kalman filter data-assimilation step
\begin{align}
x_{{\rm da},k} &= x_{{\rm fc},k} + K_{{\rm da},k} z_k, \label{kalman_da_state}
\end{align}
where the state estimator gain $K_{{\rm da},k} \in \mathbb R^{n}$, the data-assimilation error covariance $P_{{\rm da},k} \in \mathbb R^{n \times n},$
and the forecast error covariance $P_{\rmf,k+1} \in \mathbb R^{n \times n}$ are given by
\begin{align}
    K_{{\rm da},k} &= - P_{\rmf,k}C^{\rmT} ( C P_{\rmf,k} C^{\rmT} + V_{2,k}) ^{-1}, \label{kalman_gain} \\
    P_{{\rm da},k} &=  (I_{n}+K_{{\rm da},k}C) P_{\rmf,k},\label{Pda} \\
	P_{\rmf,k+1} &=  A P_{{\rm da},k}A^{\rmT} + V_{1,k}, \label{Pf}
\end{align}
$V_{1,k}\isdef B\ {\rm var}\ (d_k-\hat{d}_k)B^\rmT + A\ {\rm cov}\ (x_k - x_{{\rm da},k},d_k-\hat{d}_k)B^\rmT + B\ {\rm cov}\ (x_k - x_{{\rm da},k},d_k-\hat{d}_k)A^\rmT$, and $P_{\rmf,0} = 0.$ 

\subsection{Application of AIE to Numerical Differentiation}

To apply AIE to causal numerical differentiation, \eqref{state_eqn} and \eqref{output_eqn} are used to model a discrete-time integrator. AIE then yields an estimate $\hat{d}_k$ of the derivative of the sampled output $y_k$. 
For single discrete-time differentiation,  $A = 1, B = T_\rms,$ and $C=1$, whereas, for double discrete-time differentiation,  
\begin{align}
    A = \begin{bmatrix}
        1 & T_\rms\\ 0 & 1
    \end{bmatrix}, \quad B = \begin{bmatrix}
       \half T^2_\rms \\ {T_\rms}
    \end{bmatrix}, \quad C = \begin{bmatrix}
        1 & 0
    \end{bmatrix}.
\end{align}
%

\section{Adaptive Input and State Estimation} \label{sec:AdapInptStateEst}


In practice, $V_{1,k}$ and $V_{2,k}$ may be unknown in (\ref{Pf}) and \eqref{kalman_gain}.
To address this problem, three versions of AIE are presented.
In each version, $V_{1,k}$ and $V_{2,k}$ may or may not be adapted.
These versions are summarized in Table \ref{TableAIE}.

\begin{table}[H]
\begin{center} 
\begin{tabular}{ |c|c|c| }
 \hline
 & $V_{1,k}$ Adaptation & $V_{2,k}$ Adaptation\\
 \hline
 AIE/NSE   & No  & No \\
 \hline
 AIE/SSE   & Yes  & No \\
 \hline
 AIE/ASE   & Yes  & Yes \\
 \hline 
\end{tabular}
\end{center}
\caption{Definitions of AIE/NSE, AIE/SSE, and AIE/ASE.  Each version of AIE is determined by whether or not $V_{1,k}$ and/or $V_{2,k}$ is adapted in the state-estimation subsystem.}
\label{TableAIE}
\end{table}

To adapt ${V}_{1,k}$ and ${V}_{2,k}$, at each step $k$ we define the computable performance metric 
\begin{align}
  {J}_{k}({V}_{1,k},{V}_{2,k}) \isdef |\widehat{S}_{ k}-{S}_{ k}|, \label{J_daptmetric}
\end{align}
where $\widehat{S}_{ k}$ is the sample variance of $z_k$ over $[0,k]$ given by
\begin{align}
    \widehat{S}_{k} &= \cfrac{1}{k}\sum^{k}_{i=0}(z_i - \overline{z}_k)^2, \label{var_comp}\\
    \overline{z}_k &= \cfrac{1}{k+1}\sum^{k}_{i=0}z_i,  
\end{align}
and ${S}_{k}$ is the variance of the residual $z_k$ given by the Kalman filter, that is,
\begin{align}
{S}_{k} \isdef  C P_{{\rm f},k} C^{\rm T} + V_{2,k}.  \label{var_inno}
\end{align}
Note that \eqref{J_daptmetric} is the difference between the theoretical and empirical variances of $z_k$, which provides an indirect measure of the accuracy of $V_1$ and $V_2.$ 

\subsection{AIE with Non-adaptive State Estimation (AIE/NSE)} \label{AIE/NSE}
\begin{figure}[H]
  \begin{center}
{\includegraphics[width=0.75\linewidth]{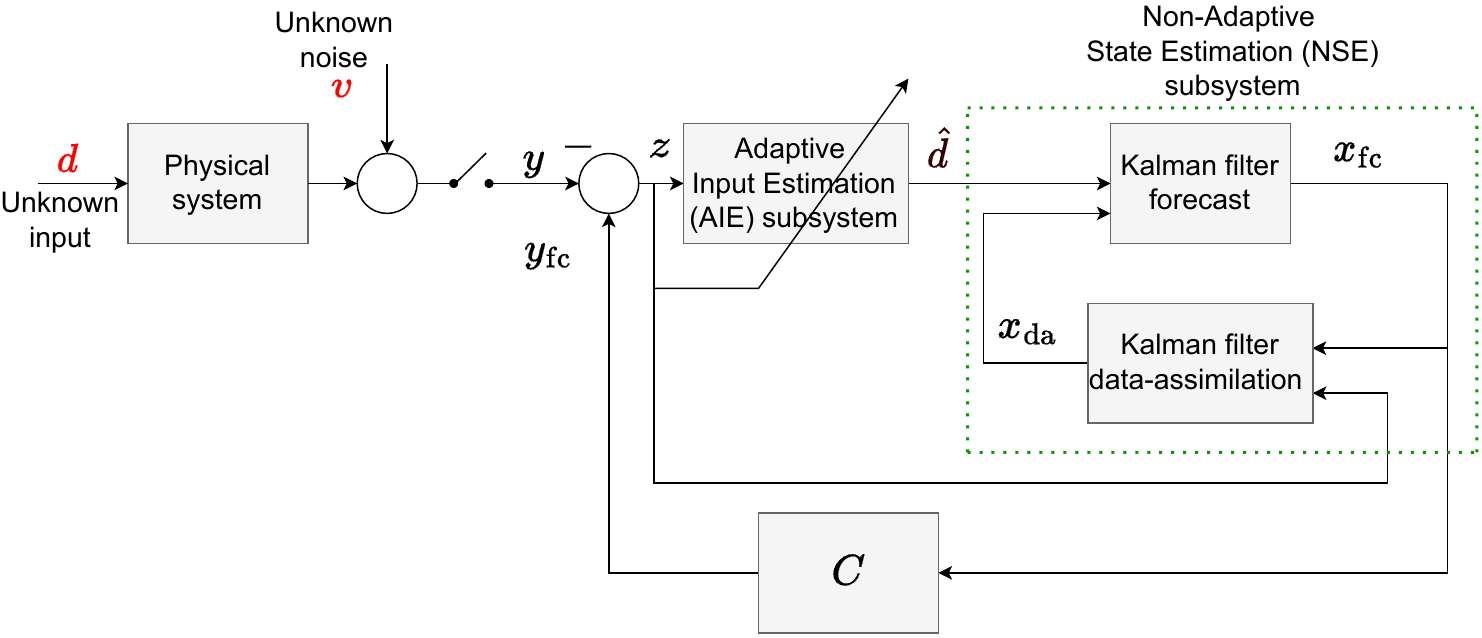}} 
\end{center}
\caption{Block diagram of AIE/NSE. 
The unknown input $d$ is the  signal whose estimates are desired, $v$ is sensor noise, and $y$ is the noisy measurement.
In this version of AIE,  $V_1$ is fixed at a user-chosen value and $V_2$ is fixed at its true value. The state estimator is thus not adaptive. }
\label{fig:sec5_block_diag_AIE_NSE}
\end{figure}

In AIE/NSE, $V_1$ is fixed at a user-chosen value, and $V_2$ is assumed to be known and fixed at its true value. 
%
%
%
%
AIE/NSE is thus a specialization of AIE with ${V}_{1,k} \equiv V_1$ in (\ref{Pf}) and  ${V}_{2,k} \equiv V_{2,{\rm true}}$ in (\ref{kalman_gain}), where $V_{2,{\rm true}}$ is the true value of the sensor-noise covariance. {A block diagram of AIE/NSE is shown in Figure \ref{fig:sec5_block_diag_AIE_NSE}.}
%

\subsection{AIE with Semi-adaptive State Estimation (AIE/SSE)} \label{AIE/SSE}

\begin{figure}[H]
  \begin{center}
{\includegraphics[width=0.75\linewidth]{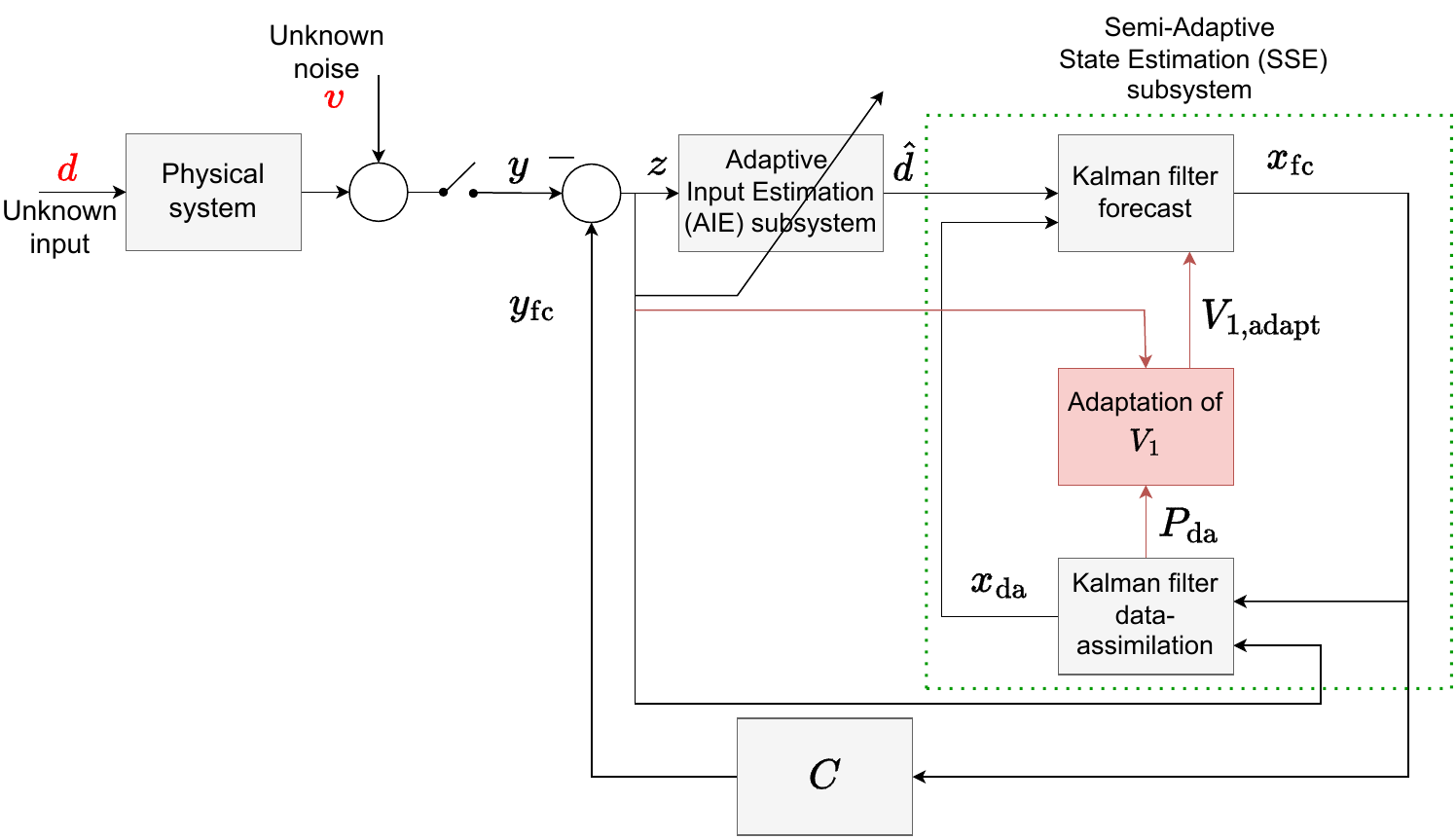}}
\end{center}
\caption{ Block diagram of AIE/SSE. In this version of AIE, $V_1$ is adapted and $V_2$ is fixed at its true value.  The state estimator is thus semi-adaptive. }
\label{fig:sec5_block_diag_AIE_SSE} 
\end{figure}

In AIE/SSE, $V_1$ is adapted, and $V_2$ is assumed to be known and fixed at its true value. 
Let ${V}_{{1,\rm adapt},k}$ denote the adapted value of $V_{1,k}$.
AIE/SSE is thus a specialization of AIE with  ${V}_{1,k} = {V}_{{1,\rm adapt},k}$ in (\ref{Pf}) and ${V}_{2,k} \equiv V_{2,{\rm true}}$ in (\ref{kalman_gain}).
%
In particular, ${V}_{{1,\rm adapt},k} = \eta I_n$ such that
\begin{align}
    {V}_{{1,\rm adapt},k} = \underset{\eta I_{n}}{\arg\min} \ J_k(\eta I_{n},V_{2,{\rm true}}), \label{V1_adpt_SSE}
\end{align}
%
where $\eta\in[\eta_{\rmL},\eta_{\rmU}]$ and 
$0 \le \eta_{\rmL} < \eta \le \eta_{\rmU}.$ {A block diagram of AIE/SSE is shown in Figure \ref{fig:sec5_block_diag_AIE_SSE}.}

%

\subsection{AIE with Adaptive State Estimation (AIE/ASE)} \label{AIE/ASE}

In AIE/ASE, both $V_1$ and $V_2$ are adapted.
Let ${V}_{{2,\rm adapt},k}$ denote the adapted value of $V_{2,k}$.
AIE/ASE is thus a specialization of AIE with ${V}_{1,k} = {V}_{{1,\rm adapt},k}$ in (\ref{Pf}) and ${V}_{2,k} = {V}_{{2,\rm adapt},k}$ in (\ref{kalman_gain}). 
In particular,  ${V}_{{1,\rm adapt},k}  = \eta I_n$ and ${V}_{{2,\rm adapt},k}\ge0$ such that 
\begin{align}
       ({V}_{{1,\rm adapt},k},{V}_{{2,\rm adapt},k}) &= \underset{ \eta I_{n},{V}_{2,k}}{\arg\min} \ J_k(\eta I_{n},V_{2,k}), \label{covmin}
\end{align} 
where $\eta\in[\eta_{\rmL},\eta_{\rmU}]$ and 
$0 \le \eta_{\rmL} < \eta \le \eta_{\rmU}.$
Defining 
\begin{align}
    {J}_{\rmf,k}(V_{1,k}) \isdef \widehat{S}_{k} - C P_{{\rm f},k}(V_{1,k})  C^{\rm T} \label{J1_func}
\end{align}
and using \eqref{var_inno}, (\ref{J_daptmetric}) can be rewritten as
\begin{align}
    {J}_k({V}_{1,k},{V}_{2,k}) = |{J}_{\rmf,k}(V_{1,k})-V_{2,k}|. \label{J_daptmetric_V2}
\end{align}
We construct a set of positive values of ${J}_{\rmf,k}$ by enumerating ${V}_{1,k} = \eta I_{n}$ as
\begin{align}
      \SJ_{\rmf,k} \isdef \{J_{\rmf,k}\colon J_{\rmf,k}(\eta I_{n}) > 0, \eta_{\rmL} \le\eta \le\eta_{\rmU}\}. \label{J_f_positive}
\end{align}
${V}_{{1,\rm adapt},k}$ and ${V}_{{2,\rm adapt},k}$ are then chosen based on following two cases.

\textbf{Case 1.}
If $\SJ_{\rmf,k}$ is not empty, then 
\begin{align}
    V_{1,{\rm adapt},k} &= \underset{\eta I_{n}}{\arg \min} \ |J_{\rmf,k}(\eta I_{n}) -  \widehat{J}_{\rmf,k}(\alpha)|,\\
  {V}_{{2,\rm adapt},k} &= J_{\rmf,k}(V_{1,{\rm adapt},k}), \label{v_2_opt_1}
\end{align}
where
\begin{align}
       \widehat{J}_{\rmf,k}(\alpha) \isdef \alpha \min \SJ_{\rmf,k}+(1-\alpha)\max \SJ_{\rmf,k}, \label{alpha1}
\end{align}
and $0 \le\alpha \le1$. For all of the examples in this paper, we set $\alpha = 1/2$ and omit the argument of $\widehat{J}_{\rmf,k}.$

\begin{figure}[H]
  \begin{center}
{\includegraphics[width=0.75\linewidth]{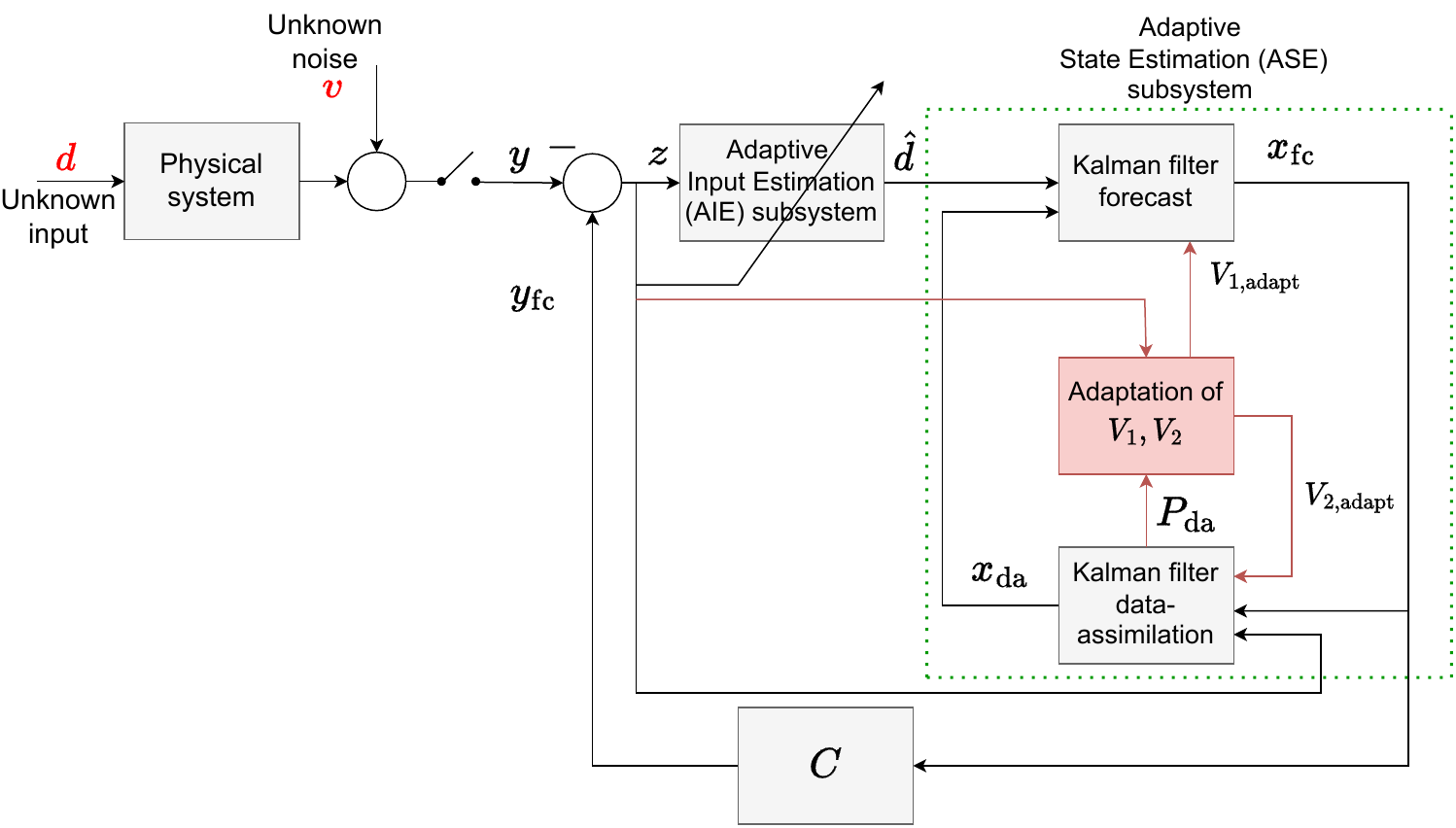}}
\end{center}
\caption{Block diagram of AIE/ASE.  In this version of AIE, both $V_1$ and $V_2$ are adapted.  The state estimator is thus adaptive. }
\label{fig:sec5_block_diag_AIE_ASE}
\end{figure}

\textbf{Case 2.}
If $\SJ_{\rmf,k}$ is empty, then
\begin{align}
    V_{1,{\rm adapt},k} &= \underset{\eta I_{n}}{\arg \min} \ |J_{\rmf,k}(\eta I_{n})|,\\
  {V}_{{2,\rm adapt},k} &= 0.  \label{v_2_opt_2}
\end{align}
{A block diagram of AIE/ASE is shown in Figure \ref{fig:sec5_block_diag_AIE_ASE}.} AIE/ASE is summarized by Algorithm \ref{AIE_ASE_algorithm}.
%


\begin{algorithm} 
\caption{Adaptive Input Estimation/Adaptive State Estimation (AIE/ASE)}\label{AIE_ASE_algorithm}

\begin{algorithmic}[1]

\State Choose $n_e \ge 1, \, n_{\rm f} \ge 1, \, R_z, \, R_d, \, R_\theta,$ $ \eta_{\rmL}, \, \eta_{\rmU}$.

\State Set $x_{{\rm fc},0} = 0$, $P_{{\rm f},0} = 0_{n\times n}$, $K_{{\rm da},0} = 0_{n\times 1}$, $\hat{d}_0 = 0$, $\theta_{k_n{-}1} = 0_{l_{\theta}\times 1}$, $P_{k_n{-}1}=R_\theta^{-1}$, $V_{1,{\rm adapt},0} = 0_{n\times n}$, $V_{2,{\rm adapt},0} = 0$.

\State $k_n = {\rm max}(n_e,n_{\rm f})$; $\widetilde{R} = {\rm blockdiag}(R_z,R_d)$;

\State \textbf{for $k=0$ to $N-1$ do}

\item[$\Comment{Residual}$]

\State\algindent $y_{{\rm fc},k} = Cx_{{\rm fc},k}$;

\State\algindent $z_{k} = y_{{\rm fc},k} - y_{k}$;

\item[$\Comment{Adaptive \ Input \ Estimation }$]

\State\algindent \textbf{if $k < k_n - 1$ do}	

\State\algindent\algindent $\hat{d}_{k} = \hat{d}_0$;

\State\algindent \textbf{else do}	

\State\algindent\algindent	$\Phi_{k} = \begin{bmatrix} \hat{d}_{k-1} & \cdots & \hat{d}_{k-n_{\rme}} & z_{k} & \cdots & z_{k-n_{\rme}} \end{bmatrix}$;

\State\algindent\algindent $\hat{d}_{k} = \Phi_{k} \theta_{k}$;

\State\algindent\algindent $\overline{A}_{k{-}1} = A (I_{n}+K_{{\rm da},k{-}1}C)$;

\State\algindent\algindent \textbf{for $i=2$ to $n_\rmf$ do}	

\State\algindent\algindent\algindent $H_{i,k} = \vspace{0.5mm} C \overline{A}_{k-1}\cdots \overline{A}_{k-(i-1)}  B$;

\State\algindent\algindent \textbf{end for}	

\State\algindent\algindent $\widetilde{H}_k = \begin{bmatrix} C B & H_{2,k} & \cdots & H_{n_{\rm f},k} \end{bmatrix}$;

\State\algindent\algindent $\Phi_{{\rm f},k} = \widetilde{H}_{k} \begin{bmatrix}
\Phi_{k{-}1}^{\rm T} & \cdots & \Phi_{k{-}n_{\rm f}}^{\rm T}
\end{bmatrix}^{\rm T} $;
\vspace{0.5mm}
\State\algindent\algindent $\hat{d}_{{\rm f},k} = \widetilde{H}_k \begin{bmatrix}
\hat{d}_{k{-}1}^{\rm T} & \cdots & \hat{d}_{k{-}n_{\rm f}}^{\rm T}
\end{bmatrix}^{\rm T} $;

\State\algindent\algindent $\widetilde{\Phi}_{k} =  \begin{bmatrix} 
   \Phi_{{\rmf},k}^{\rm T}  &  \Phi_{k}^{\rm T}
\end{bmatrix}^{\rm T}$;

\State\algindent\algindent $\widetilde{z}_{k} = \begin{bmatrix}
   (z_k-\hat{d}_{\rmf,k})^{\rm T}  & 0   \\
\end{bmatrix}^{\rm T}$;

\State\algindent\algindent $\Gamma_{k} = (\widetilde{R}^{-1} + \widetilde{\Phi}_{k} P_{k} \widetilde{\Phi}_{k}^{\rmT})^{-1}$;

\vspace{0.5mm}

\State\algindent\algindent $P_{k{+}1} =  P_{k} - P_{k}\widetilde{\Phi}_{k}^{\rmT} \Gamma_{k}\widetilde{\Phi}_{k} P_{k}$; 

\vspace{0.5mm} 

\State\algindent\algindent $\theta_{k{+}1} = \theta_{k} - P_{k} \widetilde{\Phi}_{k}^{\rmT} \Gamma_{k}(\widetilde{z}_{k} + \widetilde{\Phi}_{k} \theta_{k} )$;

\State\algindent \textbf{end if}	

\item[$\Comment{Adaptive \ State \ Estimation}$]

\State\algindent \textbf{if} $k \ge 1$ \textbf{do}
\vspace{0.5mm}

\State\algindent\algindent ${\SJ}_{\rmf,k} = [~]$; $\Comment{empty~set}$ 

\State\algindent\algindent $\widehat{S}_{k} = \texttt{variance}([z_{0} \cdots  z_k]); \Comment{ using~(\ref{var_comp})}$

\State\algindent\algindent \textbf {for $i=0$ to $w$ do}  $\Comment{ Choose ~w > 0}$

\State\algindent\algindent\algindent $\eta_i = \eta_{\rmL} + i(\eta_{\rmU}-\eta_{\rmL})/w; $
\vspace{0.5mm}

\State\algindent\algindent\algindent $\widetilde{P}_{{\rm{f}},k,i} =  A P_{{\rm{da}},k-1}A^{\rm{T}} + \eta_{i} I_{n}$;
\vspace{0.5mm}

\State\algindent\algindent\algindent $\widetilde{J}_{\rmf,k,i} = \widehat{S}_{k} - C \widetilde{P}_{{\rm{f}},k,i}  C^{\rm T}$;
\vspace{0.5mm}

\State\algindent\algindent\algindent \textbf{if} $\widetilde{J}_{\rmf,k,i} > 0$ \textbf{do}


\State\algindent\algindent\algindent\algindent  ${\SJ}_{\rmf,k} = \texttt{append}(
    {\SJ}_{\rmf,k},~\widetilde{J}_{\rmf,k,i})$;  

\State\algindent\algindent\algindent  \textbf{end if}

\State\algindent\algindent \textbf{end for}

\algstore{myalg}
  \end{algorithmic}
\end{algorithm}
\clearpage
\begin{algorithm}
  \ContinuedFloat
  \caption{Adaptive Input Estimation/Adaptive State Estimation (AIE/ASE) (continued)}
  \begin{algorithmic}
      \algrestore{myalg}

\State\algindent\algindent\textbf{if} ${\SJ}_{\rmf,k}$ is non-empty \textbf{do}
\vspace{0.5mm}

\State\algindent\algindent\algindent $\widehat{J}_{\rmf,k} =  (\min \SJ_{\rmf,k}+\max \SJ_{\rmf,k})/2$;

\State\algindent\algindent\algindent $V_{1,{\rm adapt},k} = \underset{\eta I_n}{\arg \min} \ |J_{\rmf,k}(\eta I_n) -  \widehat{J}_{\rmf,k}|$;

\State\algindent\algindent\algindent ${V}_{{2,\rm adapt},k} = J_{\rmf,k}(V_{1,{\rm adapt},k})$;

\State\algindent\algindent \textbf{else} \textbf{do}

\State\algindent\algindent\algindent $V_{1,{\rm adapt},k} = \underset{\eta I_n}{\arg \min} \ |J_{\rmf,k}(\eta I_n)|$;

\State\algindent\algindent\algindent ${V}_{{2,\rm adapt},k} = 0$;

\State\algindent\algindent \textbf{end if}

\State\algindent \textbf{end if}

\item[$\Comment{\ Kalman \ Filter \ Data-Assimilation}$]

\State\algindent $K_{\rm{da},k} =  - P_{\rm{f},k}C^{\rm{T}} (C P_{{\rm{f}},k} C^{\rm{T}} + {V}_{{2,\rm adapt},k})^{-1}$;

\State\algindent $P_{\rm{da},k} =  (I_{n} +  K_{{\rm{da}},k}C) P_{{\rm f},k}$;

\State\algindent $x_{{\rm da},k} = x_{{\rm fc},k} + K_{\rm{da},k} z_{k}$; 

\item[$\Comment{\ Kalman \ Filter \ Forecast}$]

\State\algindent $P_{{\rm{f}},k+1} =  A P_{{\rm{da}},k}A^{\rm{T}} + {V}_{{1,\rm adapt},k}$;

\State\algindent	$x_{{\rm fc},k+1} = A x_{{\rm da},k}+ B \hat{d}_{k}$

\State \textbf{end for}

\end{algorithmic}

\end{algorithm}

\section{Numerical Differentiation of Two-Tone Harmonic Signal} \label{sec:Numdiff_HS}

In this section, a numerical example is given to compare the accuracy of the numerical differentiation algorithms discussed in the previous sections. 
We consider a two-tone harmonic signal, and we compare the accuracy (relative RMSE) of BD, HGO/1, SG, AIE/NSE, AIE/SSE, and AIE/ASE. 
For single and double differentiation, the parameters for HGO/1 and SG are given in Section \ref{sec:CompDelayAlgo}.

    \begin{exam} \label{eg_sin_sin}
      {\it Differentiation of a two-tone harmonic signal}
      
    {\rm 
      Consider the continuous-time signal $y(t) = \sin(20t)+\sin(30t)$, where $t$ is time in seconds. 
      The signal $y(t)$ is sampled with sample time $T_\rms = 0.01$ sec. 
     The measurements are assumed to be corrupted by noise, and thus the noisy sampled signal is given by $y_k = \sin(0.2k)+\sin(0.3k)+D_2v_k$, where $v_k$ is standard white noise.
      %
      %

      {\it Single Differentiation.}
      For AIE/NSE, let $n_\rme = 12$, $n_\rmf = 25,$ $R_z = 1, R_d = 10^{-5}, R_\theta = 10^{-1}I_{25}, $ ${V_{1}} = 10^{-6}$, and $V_{2} = 0.01$ for SNR $20$ dB.
      For AIE/SSE, the parameters are the same as those of AIE/NSE, except that ${V_{1,k}}$ is adapted, where $\eta_{\rmL} = 10^{-6}$ and $\eta_{\rmU} = 10^2$ in Section \ref{AIE/SSE}.
      Similarly, for AIE/ASE, the parameters are the same as those of AIE/SSE except that ${V_{2,k}}$ is adapted as in Section \ref{AIE/ASE}.

      Figure \ref{fig:sec4_sin_sin_est_sd} compares the true first derivative with the estimates obtained from AIE/NSE, AIE/SSE, and AIE/ASE.
      Figure \ref{fig:sec4_sin_sin_comp_sd} shows that AIE/ASE has the best accuracy over the range of SNR.  
      Figure \ref{fig:sec4_sin_sin_v1_comp_sd} shows that the accuracy of AIE/ASE is close to the best accuracy of AIE/NSE. 

          \begin{figure}[H]
              \begin{center}
            {\includegraphics[width=0.75\linewidth]{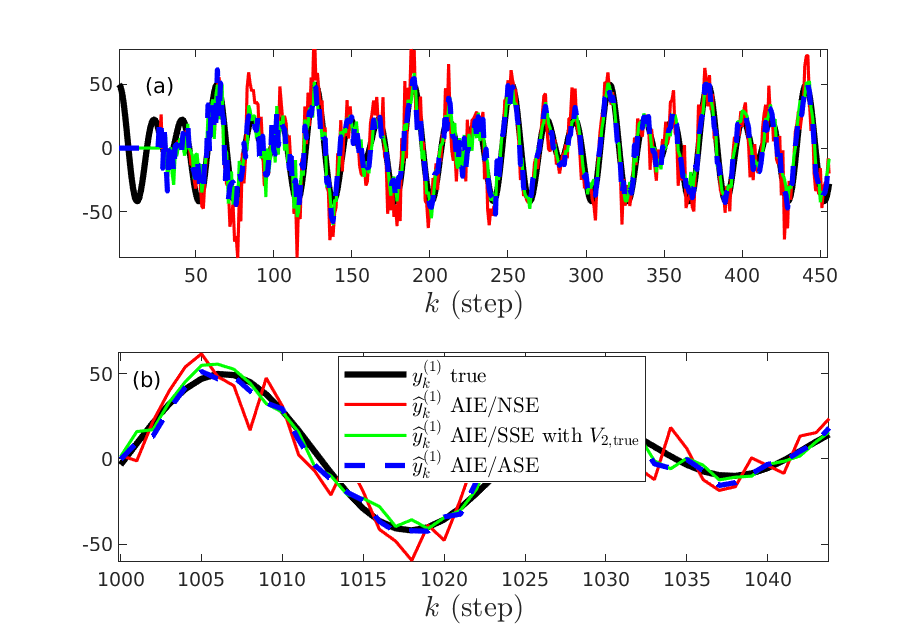}}
            \end{center}
            \caption{\textit{Example \ref{eg_sin_sin}:  Single differentiation of a sampled two-tone harmonic signal.}  (a) The numerical derivatives estimated by AIE/NSE, AIE/SSE with $V_2 = V_{2,{\rm true}}$, and AIE/ASE follow the true first derivative 
            $y^{(1)}$ after an initial transient.
            (b) Zoom of (a). At steady state, AIE/ASE is more accurate than both AIE/NSE and AIE/SSE with $V_2 = V_{2,{\rm true}}$. The SNR is 20 dB.} 
            \label{fig:sec4_sin_sin_est_sd}
          \end{figure} 
          \begin{figure}[H]
              \begin{center}
            {\includegraphics[width=0.75\linewidth]{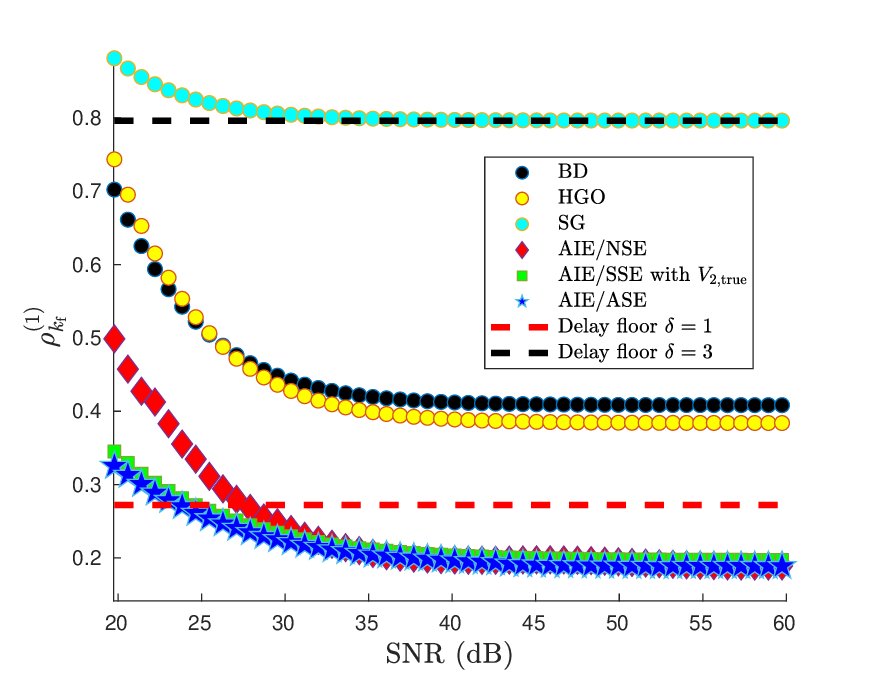}}
            \end{center}
            \caption{\textit{Example \ref{eg_sin_sin}:  Relative RMSE $\rho_{k_\rmf}^{(1)}$ of the estimate of the first derivative  of a two-tone harmonic signal
            versus SNR.} AIE/ASE has the best accuracy over the range of SNR.
            Here $k_{\rmf} = 2000$ steps.} 
            \label{fig:sec4_sin_sin_comp_sd}
          \end{figure}
          \begin{figure}[H]
            \begin{center}
          {\includegraphics[width=0.75\linewidth]{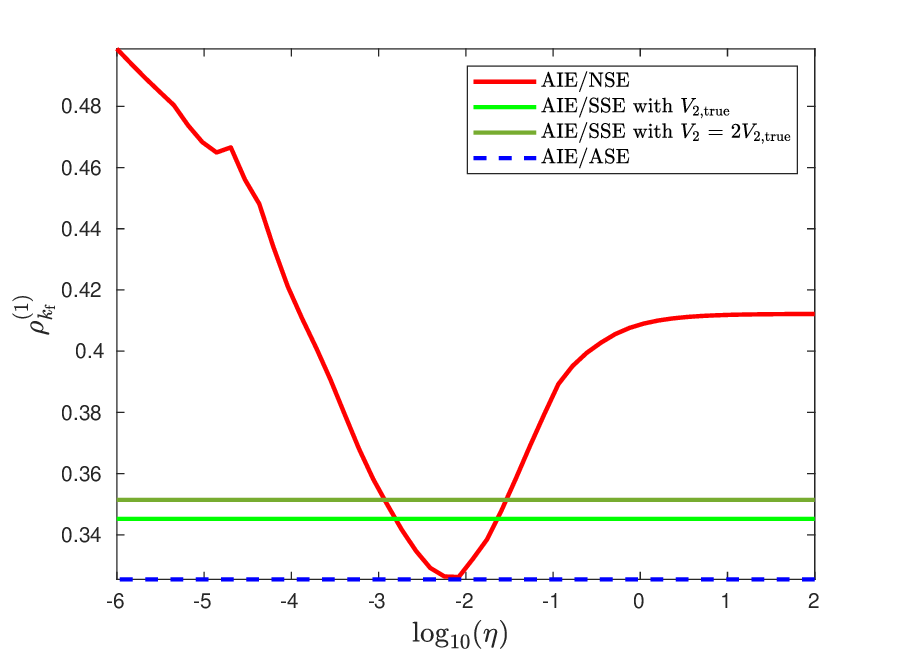}}
          \end{center}
          \caption{\textit{Example \ref{eg_sin_sin}:  Relative RMSE $\rho_{k_\rmf}^{(1)}$ of the estimate of the first derivative  of a two-tone harmonic signal versus $\eta$, such that $V_1 = \eta.$} 
          AIE/SSE with $V_2 = V_{2,{\rm true}}$ is more accurate than AIE/SSE with $V_2 = 2V_{2,{\rm true}}$, which shows the effect of $V_2$ on accuracy. 
          The accuracy of AIE/ASE is close to the best accuracy of AIE/NSE. The SNR is 20 dB, and $k_{\rmf} = 2000$ steps.} 
          \label{fig:sec4_sin_sin_v1_comp_sd}
          \end{figure}

            {\it Double Differentiation.}
          For AIE/NSE, let $n_\rme = 12$, $n_\rmf = 20,$ $R_z = 1, R_d = 10^{-5}, R_\theta = 10^{-0.1}I_{25}, $ ${V_{1}} = 10^{-1}I_2$, and $V_{2} = 0.0001$ for SNR $40$ dB.
          For AIE/SSE, the parameters are the same as those of AIE/NSE, except that ${V_{1,k}}$ is adapted, where $\eta_{\rmL} = 10^{-6}$ and $\eta_{\rmU} = 1$ in Section \ref{AIE/SSE}.
          Similarly, for AIE/ASE, the parameters are the same as those of AIE/SSE except that ${V_{2,k}}$ is adapted as in Section \ref{AIE/ASE}.

          Figure \ref{fig:sec4_sin_sin_est_dd} compares the true second derivative with the estimates obtained from AIE/NSE, AIE/SSE with  $V_2=V_{2,\rm true}$, and AIE/ASE.
          Figure \ref{fig:sec4_sin_sin_comp_dd} shows that AIE/ASE has the best accuracy over the range of SNR.  
          Figure \ref{fig:sec4_sin_sin_v1_comp_dd} shows that the accuracy of AIE/ASE is close to the best accuracy of AIE/NSE. 
              \begin{figure}[H]
                  \begin{center}
                {\includegraphics[width=0.75\linewidth]{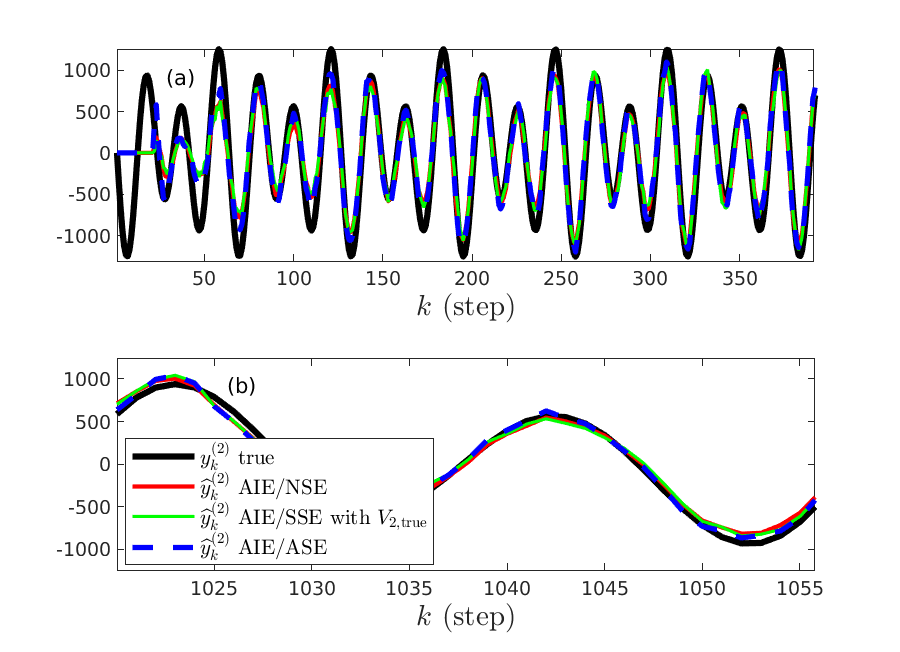}}
                \end{center}
                \caption{\textit{Example \ref{eg_sin_sin}:  Double differentiation of a sampled two-tone harmonic signal.} (a) The numerical derivatives estimated by AIE/NSE, AIE/SSE with  $V_2=V_{2,\rm true}$, and AIE/ASE follow the true second derivative 
                $y^{(2)}$ after an initial transient.
                (b) Zoom of (a). At steady state, AIE/ASE is more accurate than AIE/SSE with  $V_2=V_{2,\rm true}$ and AIE/NSE. The SNR is 40 dB.} 
                \label{fig:sec4_sin_sin_est_dd}
              \end{figure} 
              \begin{figure}[H]
                  \begin{center}
                {\includegraphics[width=0.75\linewidth]{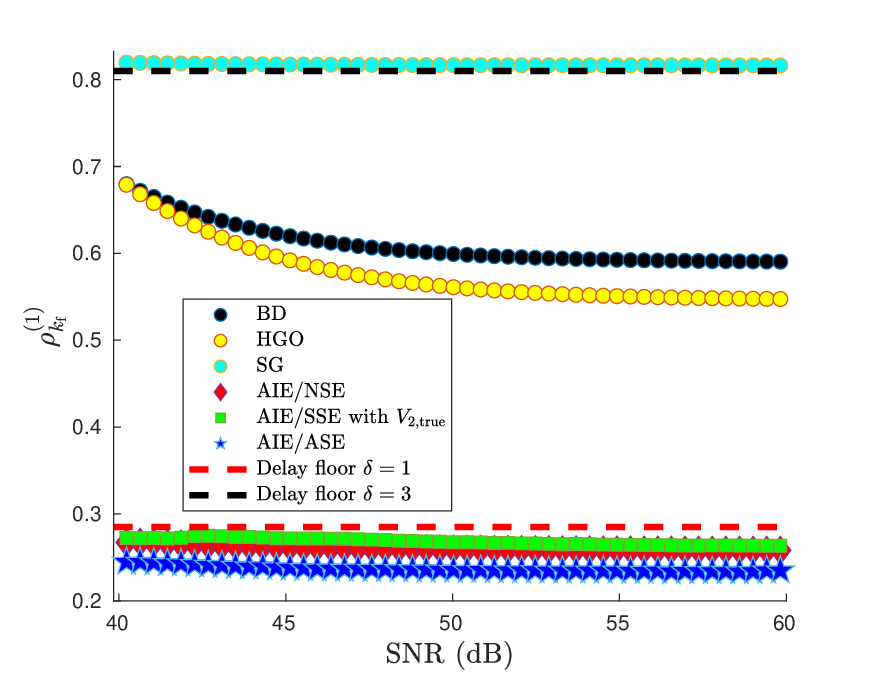}}
                \end{center}
                \caption{\textit{Example \ref{eg_sin_sin}:  Relative RMSE $\rho_{k_\rmf}^{(2)}$ of the estimate of the second derivative of a two-tone harmonic signal
                versus SNR.} AIE/ASE has the best accuracy over the range of SNR. Here $k_{\rmf} = 2000$ steps.} 
                \label{fig:sec4_sin_sin_comp_dd}
              \end{figure}
              \begin{figure}[H]
                \begin{center}
              {\includegraphics[width=0.75\linewidth]{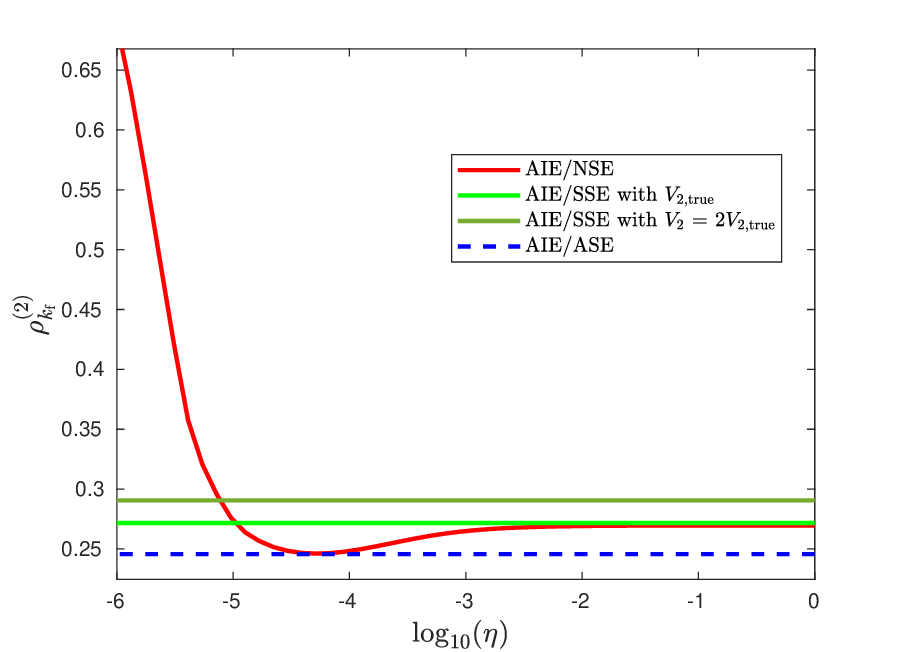}}
              \end{center}
              \caption{\textit{Example \ref{eg_sin_sin}:  Relative RMSE $\rho_{k_\rmf}^{(2)}$ of the estimate of the second derivative  of a two-tone harmonic signal versus $\eta$, such that $V_1 = \eta I_2.$} AIE/SSE with $V_2 = V_{2,{\rm true}}$
              is more accurate than AIE/SSE with $V_2 = 2V_{2,{\rm true}}$, which shows the effect of $V_2$ on accuracy. 
              The accuracy of AIE/ASE is close to the best accuracy of AIE/NSE. The SNR is 40 dB, and $k_{\rmf} = 2000$ steps.} 
              \label{fig:sec4_sin_sin_v1_comp_dd}
              \end{figure}

    }  
    \end{exam}

\section{Application to Ground-Vehicle Kinematics} \label{sec:AppCarsim}
\vspace{-0mm}

\begin{figure}[H]
  \vspace{0.2cm}
  \begin{center}
  {\includegraphics[scale = 0.4]{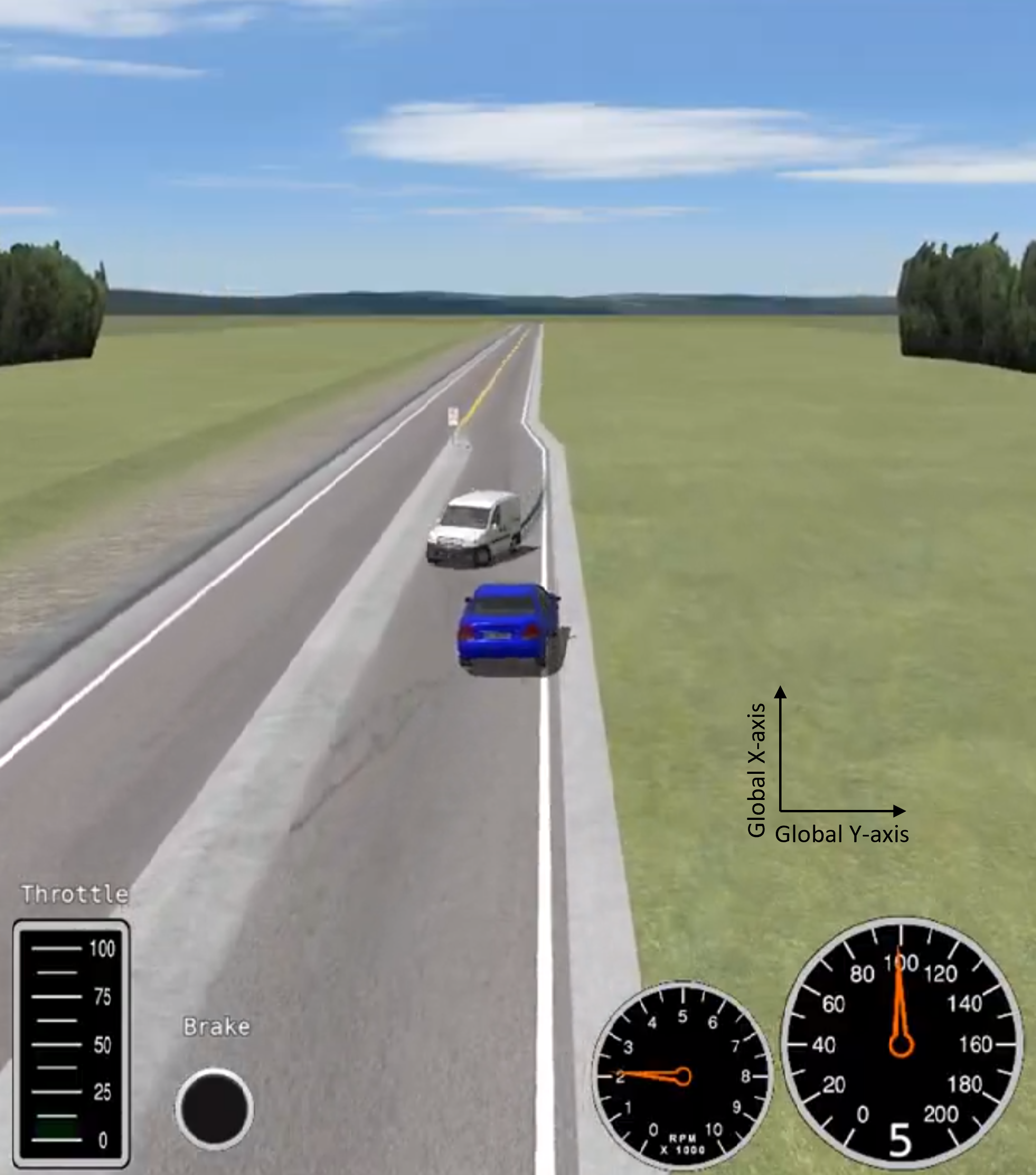}}
  \vspace{-0.0cm}
  \caption{{\it }{\it Collision-avoidance scenario in CarSim.} 
  In this scenario, the oncoming vehicle (the white van) enters the opposite lane, and the host vehicle (the blue van) performs an evasive maneuver to avoid a collision. } \label{fig_carsim}
  \end{center}
  \end{figure}

\begin{figure}[H]
  \vspace{0.2cm}
  \begin{center}
  {\includegraphics[scale = 0.6]{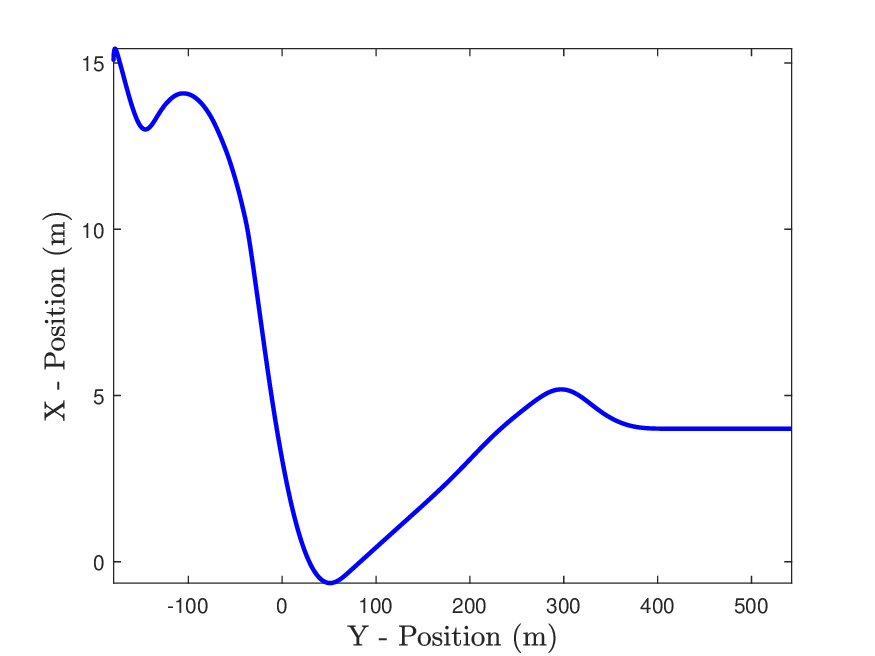}}
  \vspace{-0.0cm}
  \caption{{\it }{{\it Collision-avoidance scenario in CarSim.} 
  Relative position trajectory of the host and the target vehicles on $x$-$y$ plane.}} \label{fig_carsim_traj}
  \end{center}
  \end{figure}

In this section, CarSim is used to simulate a scenario in which an oncoming vehicle (the white van in Figure  \ref{fig_carsim}) slides over to the opposing lane. The host vehicle (the blue van) performs an evasive maneuver to avoid a collision. 
Relative position data along the global y-axis (shown in Figure  \ref{fig_carsim}) is differentiated to estimate the relative velocity and acceleration along the same axis. {Figure \ref{fig_carsim_traj} shows the relative position trajectory of the vehicles on the $x$-$y$ plane}

  \begin{exam} \label{carsim}
    {\it Differentiation of CarSim position data.}

  {\rm Discrete-time  position data generated by CarSim is corrupted with discrete-time, zero-mean, Gaussian white noise whose variance is chosen to vary the SNR.
  %
  %
  \subsubsection*{Single Differentiation}
  For AIE/NSE, let $n_\rme = 25$, $n_\rmf = 50,$ $ R_z = 1, R_d = 10^{-6}, R_\theta = 10^{-0.1}I_{51}$, ${V_{1}} = 10^{-5}$, and $V_{2} = 0.0049$ for SNR $40$ dB.
  For AIE/SSE, the parameters are the same as those of AIE/NSE, except that ${V_{1,k}}$ is adapted, where $\eta_{\rmL} = 10^{-6}$ and $\eta_{\rmU} = 10^{-2}$ in Section \ref{AIE/SSE}.
  Similarly, for AIE/ASE, the parameters are the same as those of AIE/SSE except that ${V_{2,k}}$ is adapted as in Section \ref{AIE/ASE}.

  Figure \ref{fig:sec5_carsim_rmse_40_sd} compares the true first derivative with the estimates obtained from AIE/NSE, AIE/SSE with  $V_2=V_{2,\rm true}$, and AIE/ASE. 
  Figure \ref{fig:sec5_carsim_est_40_sd} shows that the accuracy of AIE/ASE is close to the best accuracy of AIE/NSE. 
        \begin{figure}[H]
          \begin{center}
        {\includegraphics[width=0.75\linewidth]{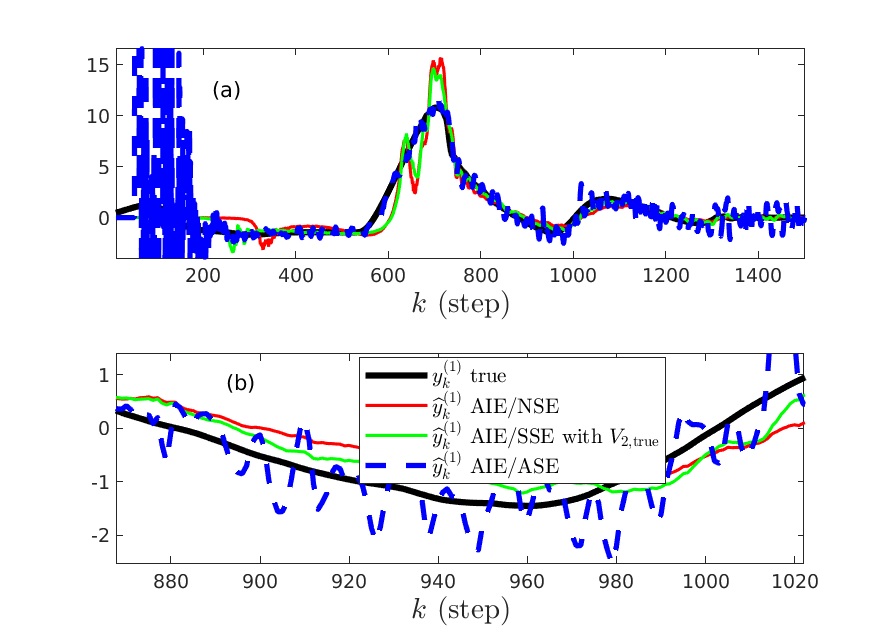}}
        \end{center}
        \caption{\textit{Example \ref{carsim}:  Single differentiation of CarSim data.}  (a) The numerical derivatives estimated by AIE/NSE, AIE/SSE with  $V_2=V_{2,\rm true}$, and AIE/ASE follow the true first derivative $y^{(1)}$ after an initial transient of 200 steps.
        (b) Zoom of (a). At steady state, AIE/ASE is more accurate than both AIE/NSE and AIE/SSE with  $V_2=V_{2,\rm true}$. The SNR is 40 dB.} 
        \label{fig:sec5_carsim_rmse_40_sd}
      \end{figure}

      \begin{figure}[H]
            \begin{center}
          {\includegraphics[width=0.75\linewidth]{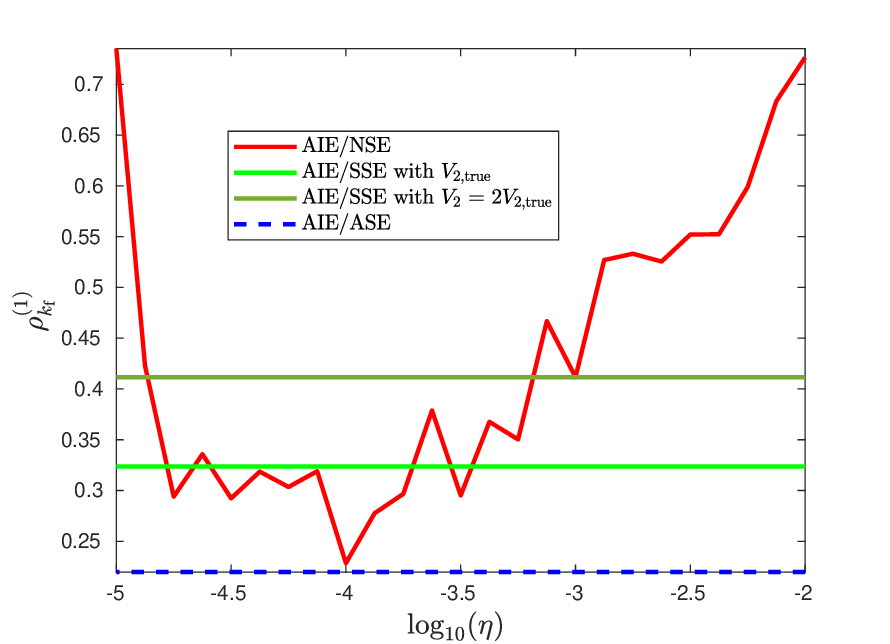}}
          \end{center}
          \caption{\textit{Example \ref{carsim}: Relative RMSE $\rho_{k_\rmf}^{(1)}$ of the estimate of the first derivative of CarSim data versus $\eta$, such that $V_1 = \eta.$} AIE/SSE with $V_2 = V_{2,{\rm true}}$ is more accurate than AIE/SSE with $V_2 = 2V_{2,{\rm true}}$, which shows the effect of $V_2$ on accuracy. The accuracy of AIE/ASE is close to the best accuracy of AIE/NSE. 
            The SNR is 40 dB, and $k_{\rmf} = 1500$ steps.} 
          \label{fig:sec5_carsim_est_40_sd}
      \end{figure}

      \subsubsection*{Double Differentiation}
      For AIE/NSE, Let $n_\rme = 25$, $n_\rmf = 21,$ $R_z = 1, R_d = 10^{-5}, R_\theta = 10^{-8}I_{51},$ ${V_{1}} = 10^{-3}I_2$, and $V_{2} = 0.0049$ for SNR $40$ dB.
      For AIE/SSE, the parameters are the same as those of AIE/NSE, except that ${V_{1,k}}$ is adapted, where $\eta_{\rmL} = 10^{-3}$ and $\eta_{\rmU} = 1$ in Section \ref{AIE/SSE}.
      Similarly, for AIE/ASE, the parameters are the same as those of AIE/SSE except that ${V_{2,k}}$ is adapted as in Section \ref{AIE/ASE}.

      Figure \ref{fig:sec5_carsim_rmse_40_dd} compares the true second derivative with the estimates obtained from AIE/NSE, AIE/SSE with  $V_2=V_{2,\rm true}$, and AIE/ASE.
      Figure \ref{fig:sec5_carsim_est_40_dd} shows that the accuracy of AIE/ASE is close to the best accuracy of AIE/NSE. 
      \begin{figure}[H]
        \begin{center}
      {\includegraphics[width=0.75\linewidth]{carsim_estimate_dd.eps}}
      \end{center}
      \caption{\textit{Example \ref{carsim}:  Double differentiation of CarSim data.}  (a) The numerical derivatives estimated by AIE/NSE, AIE/SSE with $V_2=V_{2,\rm true}$, and AIE/ASE follow the true first derivative $y^{(2)}$ after an initial transient.
      (b) Zoom of (a). At steady state, AIE/ASE is more accurate than both AIE/NSE and AIE/SSE with $V_2=V_{2,\rm true}$. The SNR is 40 dB.} 
      \label{fig:sec5_carsim_rmse_40_dd}
    \end{figure}
    \begin{figure}[H]
          \begin{center}
        {\includegraphics[width=0.75\linewidth]{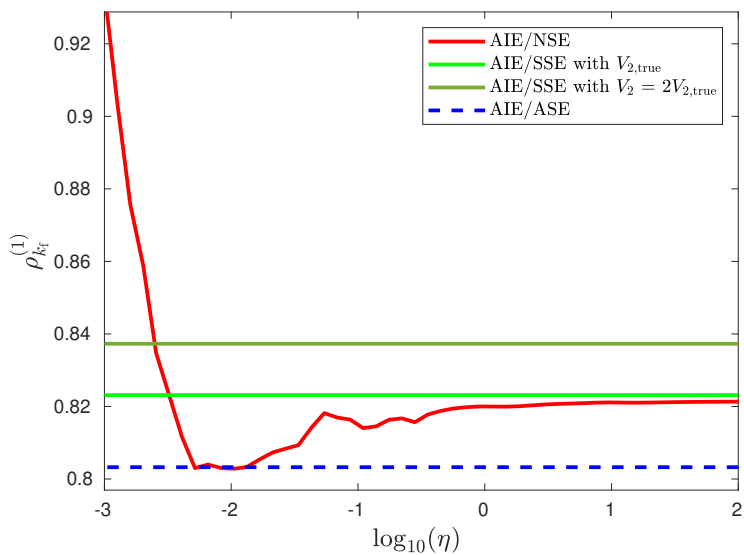}}
        \end{center}
        \caption{\textit{Example \ref{carsim}:  Relative RMSE $\rho_{k_\rmf}^{(2)}$ of the estimate of the second derivative of CarSim data versus $\eta$, such that $V_1 = \eta I_2.$} 
        AIE/SSE with $V_2 = V_{2,{\rm true}}$ is more accurate than AIE/SSE with $V_2 = 2V_{2,{\rm true}}$, which shows the effect of $V_2$ on accuracy. 
        The accuracy of AIE/ASE is close to the best accuracy of AIE/NSE.  The SNR is 40 dB, and $k_{\rmf} = 1500$ steps.} 
        \label{fig:sec5_carsim_est_40_dd}
    \end{figure}   
} 
\end{exam}

\section{Conclusions}

This paper presented the adaptive input and state estimation algorithm AIE/ASE for causal numerical differentiation.
AIE/ASE uses the Kalman-filter residual to adapt the input-estimation subsystem and an empirical estimate of the estimation error to adapt the input-estimation and sensor-noise
covariances.
For dual-tone harmonic signals with various levels of sensor noise, the accuracy of AIE/ASE was compared to several conventional numerical differentiation methods.
Finally, AIE/ASE was applied to simulated vehicle position data generated by CarSim.

Future work will focus on the following extensions.
The minimization of \eqref{covmin} was performed by using a gridding procedure;  more efficient optimization is possible.
Next, the choice of $\alpha = 1/2$ in \eqref{alpha1} was found to provide good accuracy in all examples considered;  however, further investigation is needed to refine this choice.
Furthermore, it is of interest to compare the accuracy of AIE/ASE to the adaptive sliding mode differentiator in \citet{alwi_adap_sliding_mode_2012}.
Finally, in practice, the spectrum of the measured signal and sensor noise may change abruptly.
In these cases, it may be advantageous to replace the RLS update \eqref{covariance_update}, \eqref{theta_update} with RLS that uses variable-rate forgetting in \citet{adamVRF,nimaFtest}.

\section*{Acknowledgments}
This research was supported by Ford and NSF grant CMMI 2031333.

\section*{Disclosure statement}
No potential conflict of interest was reported by the author(s).

%
\bibliographystyle{apacite}
\bibliography{bibliography}
\vspace{-0.5cm}
\end{document}